\documentclass[onecolumn,preprintnumbers,amsmath,amssymb]{revtex4}

\usepackage{graphicx}
\usepackage{dcolumn}
\usepackage{bm}

\begin{document}


\title{Regularized braneworlds of arbitrary codimension}

\author{Stephen A. Appleby and Richard A. Battye \\ {\it Jodrell Bank
Observatory, Department of Physics and Astronomy,} \\ {\it
University of Manchester, Macclesfield, Cheshire, SK11 9DL}}


\date{\today}

\begin{abstract}

We consider a thick $p$-brane embedded in an $n$-dimensional
spacetime possessing radial symmetry in the directions orthogonal to the brane. We first consider a static brane, and find a general
fine tuning relationship between the brane and bulk parameters
required for the brane to be flat. We then consider the cosmology of
a time dependent brane in a static bulk, and find the Friedmann
equation for the brane scale factor $a(t)$. The singularities that would ordinarily arise when considering arbitrary codimensions are avoided by regularizing the brane, giving it a finite profile in the transverse dimensions. However, since we consider the brane to be a strictly local defect, we find that the transverse dimensions must have infinite volume, and hence gravity cannot be localized on the brane without resorting to some infra-red cutoff.

\end{abstract}

\maketitle

\section{Introduction}

The idea that our Universe might be a four dimensional subspace
embedded in a higher dimensional manifold has been exhaustively
studied over the past decade; see for example \cite{br03,shi1,rg1,ma04,rd04,cz100,rs01,rs02,cl1,ct102,dl10} and references therein (and previously;
see \cite{vr00}, \cite{vr01} and \cite{bc00}). The most
prominent model to have arisen since the inception of so called
`braneworld' theories is the Randall Sundrum (RS) model. In Refs.
\cite{rs01}, \cite{rs02}, a thin, static, pure tension brane is
embedded in five dimensional Anti-de-Sitter (AdS) space. It was found that
a flat hypersurface could be embedded into the background space, but
only if the brane tension was fine tuned to the bulk cosmological
constant. Moreover, it was found that gravity could be localized in
the vicinity of the four dimensional static brane. This important
result meant that observers on a brane would feel standard four
dimensional gravity at sufficiently low energies, where any massive
bulk modes would be highly suppressed. In addition, this model
offered a possible solution to the hierarchy problem.

Simple codimension one braneworld models such as the Randall Sundrum
setup typically involve a static brane in a static higher
dimensional space. However, the cosmological generalization to a
time dependent brane has been considered by many authors since (see Ref.
\cite{dl10} and references therein, for example). For codimension
one objects, modifications to the standard four dimensional
Friedmann equation have been found; specifically, one obtains a new
radiation-like term that can be considered as a contribution to the
brane evolution due to the bulk. In addition, one finds another
modification to standard four dimensional cosmology which is
quadratic in the brane energy density. This implies that for large
energy densities on the brane, one would experience a completely
different brane evolution than would be the case in conventional
four dimensional gravity.

Progress in determining the cosmological evolution of codimension
two branes has also been made \cite{n01,cl00,jv00,pb00}. Initially
seen as promising candidates for solving the cosmological constant
problem, such models run into difficulties when one attempts to
introduce matter on an otherwise pure tension brane. Specifically
one obtains a singularity at the position of the brane, when
considering more general equations of state for matter localized on
the source. However, by treating the introduction of matter as a
perturbation around a known static six dimensional braneworld model,
it has been found in Ref. \cite{jv00} that if a thick brane is
considered, then we can obtain standard late time FRW behaviour.
Thickening the brane is one method of regularizing codimension two
objects, but there are others (introducing Gauss-Bonnet terms for
example \cite{pb00}).

Branes of arbitrary codimension have not been studied in general
(although see for example Refs. \cite{is1,cz20,lb1}), for a number
of reasons. In particular, one cannot simply embed a thin $(3+1)$-brane
in an $n$-dimensional space, since one generically finds naked
singularities in curvature invariants at the position of the brane. This problem can be
resolved in a number of ways; for example one could thicken the
brane, effectively smearing any singularities over a small region of
the transverse space. This is the approach that we will take, that is we will replace the singular brane energy momentum tensor with some smoothed distribution. We then define effective four dimensional quantities by integrating the full field equations over the codimensions, averaged by the brane profile. Our approach follows Refs. \cite{ct102,ct103,ct104}, where a similar regularization was considered.

This paper will be concerned with the cosmological evolution of a
thick $(3+1)$-brane of arbitrary codimension $m$. We
will show that four dimensional brane cosmology can be obtained for
a large class of metrics where a thick $(3+1)$-brane is embedded in
an $n$-dimensional background space. Our work is an extension of
Refs. \cite{n01} and \cite{n02}, where the cosmological equations for
the brane scale factor $a(t)$ were obtained for codimension one and
two branes. In Ref. \cite{n01} it was found that conventional late
time cosmology can be obtained on a four dimensional brane in a six
dimensional space. Our calculation closely follows these works, and
we find a corresponding set of equations for a brane of arbitrary
codimension. The method used in Ref. \cite{n01} to derive the
cosmological equations will generalize for higher codimension, and we will show
that we can recover standard late time cosmology regardless of the
number of extra dimensions, subject to the assumptions we impose.
We will concentrate on the case of a highly symmetric bulk spacetime, in which the brane is static. Some authors \cite{uz01,sas01,cc06} have considered dropping exact spherical symmetry in the bulk and studying the resulting brane field equations. This scenario is considerably more complicated than the models considered here, since the brane will generically experience a force when the bulk is not symmetric, and we will not consider such behaviour.

One important aspect of braneworld models is the requirement that conventional four dimensional gravity must be obtainable on the brane. For this to occur, we must have some mechanism that localizes gravity in the vicinity of the brane. One way in which this can be achieved is to have a transverse space with finite volume, as in the RS model. However, in this paper we will be considering local defects of arbitrary codimension, and it has been found in Ref. \cite{gh00} that for such models the transverse dimensions must have infinite volume. This means that gravity cannot be localized on the brane without introducing some large distance cutoff when we integrate over the transverse dimensions to obtain four dimensional gravity. This cutoff could arise by introducing a second brane, for example. We will consider this issue in section \ref{sec:100}.

The paper will proceed as follows. In the following section we
review some important definitions and general braneworld results
that we will need for the rest of the paper. In section
$\ref{sec:st}$, we find the field equations for a particular class
of static, spherically symmetric metrics. We then introduce a
time-dependent metric in section $\ref{sec:1}$ and obtain an
evolution equation for the brane scale factor, and show that our
result correctly reduces to the codimension one and two examples as
found in Refs. \cite{n01} and \cite{n02}. We write our result as a
modified four dimensional Friedman equation, and doing so we find a
non-standard cosmological equation with a dark radiation term
present. The origin of this dark radiation term is explored in
the appendix.

\section{Formalism}

We begin with some important results that will be required in
subsequent sections. In this paper, we will be considering
codimension $m$, $p$-branes ($p$ being the number of spacetime
dimensions of the brane) embedded in $n$-dimensional background
spacetimes, hence $m=n-p$. Specifically, we will consider two
different metrics. The first is a static, $n$-dimensional metric
with spherical symmetry in the extra dimensions. It takes the form

\begin{equation}\label{eq:1} ds^{2} = f(r)g_{AB}(x)dx^{A}dx^{B} -
dr^{2} - \alpha^{2}(r) \gamma_{ab}(y)dy^{a}dy^{b} ,\end{equation}

\noindent where $f(r)$ and $\alpha(r)$ are functions of the radial
coordinate $r$ only, and $\gamma_{ab}$ is the metric of the unit
$(m-1)$ sphere, so the extra dimensions are radially symmetric. The
metric $g_{AB}(x)$ is the $(3+1)$ dimensional brane metric, and
hence capital Latin indices $(A,B,..)$ run over the standard $(3+1)$
coordinates. The metric $\gamma_{ab}$ is that of the $(m-1)$-sphere
(that is all of the brane orthogonal coordinates except the radial
coordinate $r$), and hence lower case Latin indices $(a,b,..)$ run
over $(m-1)$ of the codimensions. Finally, we will use the notation
that Greek indices $(\mu,\nu,..)$ run over all $n$-dimensional
coordinates. The full $n$-dimensional metric is $G_{\mu\nu}$, and in
subsequent sections we will frequently need the measure $\sqrt{G}$,
which is given by

\begin{equation} \sqrt{G} = \lbrack f(r)\rbrack^{2} \alpha^{m-1}\sqrt{g\gamma} , \end{equation}

\noindent where $g$ and $\gamma$ are the determinants of $g_{AB}$
and $\gamma_{ab}$ respectively.

In the next section, we will need the following decompositions of
the $n$-dimensional Ricci tensor $R_{\mu\nu}$

\begin{equation} \label{eq:2} R_{AB} = R^{(g)}_{AB} -
{1 \over 2}g_{AB} (\nabla_{a}\nabla^{a} + \nabla_{r}\nabla^{r}) f(r)
- {p -2 \over 4}g_{AB} f(r)^{-1} \nabla_{r}f(r)  \nabla^{r}f(r)
,\end{equation}

\begin{equation} \label{eq:3} R_{ab} = R^{(\gamma)}_{ab} - {p \over
2}f^{-1} \nabla_{a}\nabla_{b} f  + {p \over 4}f^{-2}
\left(\nabla_{a}f\right) \left(\nabla_{b}f \right),
\end{equation}

\begin{equation} \label{eq:4} R_{rr} = R^{(\gamma)}_{rr} - {p \over
2}f^{-1} \nabla_{r}\nabla_{r} f  + {p \over 4}f^{-2}
\left(\nabla_{r}f\right) \left(\nabla_{r}f \right),
\end{equation}

\noindent where we will take $p=4$ in this paper, but is
in general $p = g_{AB}g^{AB}$, and $R^{(g)}_{AB}$ is the Ricci
tensor constructed from $g_{AB}(x)$. $R_{ab}^{(\gamma)}$ and
$R_{rr}^{(\gamma)}$ are the components of the Ricci tensor
constructed from the metric

\begin{equation} ds^{2}_{[{\rm m}]} = -dr^{2} - \alpha(r)^{2}
\gamma_{ab}(y)dy^{a}dy^{b} .\end{equation}

In the next section, we will find that the Einstein equations for
our metric ($\ref{eq:1}$) can be written in a very simple form in
terms of the extrinsic curvatures of the subspaces with metrics
$g_{AB}$ and $\gamma_{ab}$. For objects of codimension
greater than one the extrinsic curvature is defined as \cite{c1}

\begin{equation}\label{eq:extr1} K_{\mu\nu}{}^{\rho} \equiv \eta_{\nu}{}^{\sigma} \eta_{\mu}{}^{\beta} \nabla_{\beta}
\eta_{\sigma}{}^{\rho} ,\end{equation}

\noindent where $\nabla_{\beta}$ preserves the full
$n$-dimensional background metric, and $\eta_{\nu}{}^{\sigma}$ projects tensors
tangentially to the brane. For our metric, $\eta_{\nu}{}^{\sigma}$
is given by

\begin{equation} \eta_{\mu}{}^{\nu} =  \left( \begin{array}{cc}
g_{A}{}^{B} & 0  \\
0 & 0  \\
 \end{array} \right) .\end{equation}

\noindent The first two indices of the extrinsic curvature tensor are tangential to the brane, and the last is orthogonal. The metrics that we are considering are
highly symmetric, and as a result the extrinsic curvature tensor
simplifies considerably. Using the metric ($\ref{eq:1}$), the only non-zero
components of ($\ref{eq:extr1}$) are for $\rho = r$. We will use the extrinsic curvatures for our metric, $K_{A}{}^{B r}$ and $K_{a}{}^{b
r}$, which are given by

\begin{equation} K_{A}{}^{B r} \equiv K_{A}{}^{B} = {f' \over
f}g_{A}{}^{B}, \qquad \qquad \qquad  K_{a}{}^{b r} \equiv L_{a}{}^{b}
= 2{\alpha' \over \alpha}\gamma_{a}{}^{b} ,\end{equation}

\noindent where primes denote derivatives with respect to $r$. We
have dropped the third index on the extrinsic curvature tensor,
since the only non zero components of ($\ref{eq:extr1}$) are for
$\rho =r$.

The second metric ansatz that we will consider is

\begin{equation}\label{eq:197} ds^{2} = N^{2}(r,t) dt^{2} -
A^{2}(r,t)g_{ij}(x)dx^{i}dx^{j} - dr^{2} - \alpha^{2}(r,t)
\gamma_{ab}(y) dy^{a}dy^{b} ,\end{equation}

\noindent which is a more general version of ($\ref{eq:1}$), and will be used to
model a time dependent brane. As before, we can split the line element ($\ref{eq:197}$) into brane
tangential and brane orthogonal components. The `brane' line element,
$ds^{2}_{[b]}$, is given by

\begin{equation} ds_{[{\rm b}]}^{2} = N^{2}(r,t) dt^{2} -
A^{2}(r,t)g_{ij}(x)dx^{i}dx^{j} ,\end{equation}

\noindent and for surfaces of constant $r$ is of the form of an FRW metric.
We have introduced another set of indices $(i,j)$ in
($\ref{eq:197}$), which run over the standard three spatial
dimensions. For the metric ($\ref{eq:197}$), the measure $\sqrt{G}$
is given by

\begin{equation} \sqrt{G} = N A^{3} \alpha^{m-1} \sqrt{\gamma} .
\end{equation}

\noindent The extrinsic curvatures $K_{A}{}^{B}$ and $L_{a}{}^{b}$ for this metric are

\begin{equation}\label{eq:si1}   K_{i}{}^{j} = 2\delta_{i}{}^{j} {A'\over A} , \qquad
\qquad \qquad  K_{t}{}^{t} = 2 {N' \over N} , \qquad \qquad \qquad
L_{a}{}^{b} = 2 \delta_{a}{}^{b} {\alpha' \over \alpha},
\end{equation}

\noindent and taking the trace gives

\begin{equation} K = 2{N' \over N} + 6 {A' \over A},  \qquad \qquad \qquad L =
2(m-1){\alpha' \over \alpha} . \end{equation}

\subsection{Regularization of the brane}

Finally in this section, we explain our method of regularizing higher codimension branes, which follows Refs. \cite{ct102,ct103,ct104,uz01}. For a thin brane, the total energy momentum tensor, $T_{\mu\nu}$, can be
decomposed into distinct brane and bulk components. In the thin case, the
standard definition of the $p$-brane energy momentum tensor
$\hat{T}_{\mu\nu}$ is

\begin{equation}\label{eq:397} \hat{T}_{\mu\nu} = \int \sqrt{g}
d^{p}\sigma
\bar{T}_{\mu\nu}\delta^{(n)}\left[x_{\alpha}-X_{\alpha}(\sigma^{A})\right],
\end{equation}

\noindent where $\bar{T}_{\mu\nu}$ is the brane supported energy
momentum tensor, $\sigma^{A}$ are the brane coordinates, and
$x_{\alpha}$ the $n$-dimensional background coordinates. The brane is situated at $x_{\alpha} = X_{\alpha}(\sigma^{A})$.

In this paper we are considering branes of finite thickness. This means that we no longer treat the brane energy momentum tensor $\hat{T}_{\mu}{}^{\nu}$ as a singular source as in ($\ref{eq:397}$), but rather as some smoothed distribution. Explicitly, we consider the energy momentum tensor

\begin{equation} \label{eq:fg1} \hat{T}_{\mu\nu} = \int \sqrt{g} d^{p}\sigma \bar{T}_{\mu\nu} D^{(n)}_{\epsilon}(x- X(\sigma)) ,\end{equation}

\noindent where we have replaced the $n$-dimensional $\delta$-function in ($\ref{eq:397}$) with the finite brane profile function $D^{(n)}_{\epsilon}(x- X(\sigma))$, which is peaked at $x=X(\sigma)$ and falls away sharply from the brane.

In this paper, we will consider the particular simple brane profile

\begin{align}\label{eq:mn1} D_{\epsilon}^{(n)} = \quad &1 \quad r < \epsilon \\
  &0 \quad r > \epsilon ,\end{align}

\noindent where $\epsilon$ can be considered as the brane thickness parameter. This profile has been considered previously, see for example Refs. \cite{dl10,n02} (and a different brane profile was considered in \cite{uz01}). Although we use this particular profile, we expect that our results will be approximately valid for a large class of profile functions. Using ($\ref{eq:mn1}$), the brane supported energy momentum tensor $\tilde{T}_{\mu\nu}$ is given by

\begin{equation} \tilde{T}_{\mu}{}^{\nu} = {1 \over \sqrt{g}|_{r=\epsilon}}
\int d^{m-1}y  \int_{0}^{\epsilon} \sqrt{G} dr T_{\mu}{}^{\nu} .
\end{equation}

We must also define the bulk energy momentum tensor, $T_{\mu\nu}^{\rm bulk}$. For simplicity, we will consider a cosmological constant only in the bulk, with no
additional fields, so $T_{\mu\nu}|^{\rm bulk} = -\Lambda g_{\mu\nu}$, where $\Lambda$ is the $n$-dimensional cosmological
constant.

Finally, we must discuss how to obtain four-dimensional equations for a thick brane. For thin branes, effective four dimensional equations are found by taking the full $n$-dimensional field equations and evaluating them on a surface of constant $y_{a}$ (the codimensions), at the position of the brane. This is equivalent to taking the full $n$-dimensional equations, and integrating them over the $m$ codimensions, weighted by the brane profile, which in this case is a delta function. For thick branes, we follow the same procedure; to obtain four-dimensional equations, we take the full $n$-dimensional equations, integrate them over the codimensions, weighted by the brane profile $D_{\epsilon}^{(n)}$, which is no longer singular. For the profile ($\ref{eq:mn1}$) used in this paper, our approach corresponds to integrating the field equations over the range $r=(0,\epsilon)$. In the next section, we will perform our method of regularization for a codimension $m$ static brane.

\section{\label{sec:st} Static braneworld model}

Let us first consider a braneworld model with a metric of the form
($\ref{eq:1}$). This is the simplest generalization of codimension
one and two braneworlds that exist in the literature. Our approach follows the work of Refs. \cite{n01,n02}; we will find that all of the field
equations for the metric ansatz ($\ref{eq:1}$), except the $(r,r)$
equation, can be written approximately as total derivatives with respect to $r$. These equations can be integrated over the brane thickness to
obtain a set of junction conditions. We then evaluate the remaining
$(r,r)$ field equation at the brane-bulk boundary, that is at
$r=\epsilon$, and use our junction conditions to write an equation
relating the brane Ricci scalar $R^{(g)}$ to the brane energy momentum
tensor $\tilde{T}_{\mu\nu}$. Before continuing we note that static metrics of the form ($\ref{eq:1}$) have been considered previously, see for example \cite{gh00}.

To begin, we write the decompositions of the background Ricci tensor
$R_{\mu}{}^{\nu}$, ($\ref{eq:2}$-$\ref{eq:4}$), in terms of the
extrinsic curvature tensors $K_{A}{}^{B}$ and $L_{a}{}^{b}$,

\begin{equation}\label{eq:b6} \sqrt{G} R_{A}{}^{B} = {\sqrt{G} \over
f}R^{(g)}{}_{A}{}^{B} + {1 \over 2}\left(\sqrt{G}K_{A}{}^{B}\right)'
= {\sqrt{G} \over M^{n-2}}\left( T_{A}{}^{B} - \delta_{A}{}^{B}{T
\over n-2}\right) ,
\end{equation}

\begin{equation}\label{eq:b7} \sqrt{G}R_{a}{}^{b} = {\sqrt{G} \over
\alpha^{2}} (m-2)\delta_{a}{}^{b} +{1 \over 2}
\left(\sqrt{G}L_{a}{}^{b}\right)' = {\sqrt{G} \over M^{n-2}}\left(
T_{a}{}^{b} - \delta_{a}{}^{b}{T \over n-2}\right) ,\end{equation}

\begin{equation}\label{eq:b8} \sqrt{G} R_{r}{}^{r} =\sqrt{G} \left({1
\over 2} \left( L' + K'\right) + {1 \over
4}\left(K_{C}{}^{D}K_{D}{}^{C} + L_{a}{}^{b}L_{a}{}^{b}\right)
\right) = {\sqrt{G} \over M^{n-2}}\left( T_{r}{}^{r} - {T \over
n-2}\right) ,\end{equation}

\noindent where primes denote derivatives with respect to the radial
coordinate $r$. $M$ is the $n$-dimensional fundamental mass scale.
Note that equation ($\ref{eq:b7}$) does not exist for codimension one branes; we will
return to this specific case shortly.

We now assume that derivatives tangential to the brane (that is
derivatives with respect to the $x_{A}$ coordinates) will be
negligible compared to derivatives in the brane-orthogonal
directions. This implies that we can neglect the term
$R^{(g)}{}_{A}{}^{B} / f$ in ($\ref{eq:b6}$), which then becomes

\begin{equation}\label{eq:b90} \sqrt{G}R_{A}{}^{B} \approx  {1 \over
2}\left(\sqrt{G}K_{A}{}^{B}\right)' = {\sqrt{G} \over M^{n-2}}\left(
T_{A}{}^{B} - \delta_{A}{}^{B}{T \over n-2}\right)
\end{equation}

\noindent and we see that the ($A,B$) components of the Ricci tensor
can be approximately written as a total derivative with respect to
$r$. The next step is to integrate this expression over the brane thickness $0<r<\epsilon$. The integral of $(\sqrt{G}K_{A}{}^{B})'$ over
the brane thickness will give us two terms, one at $r=0$ and one at
$r= \epsilon$. We must choose our boundary conditions at $r=0$
carefully so that the metric is regular there, since a poor choice
will generally create naked singularities in curvature invariants at
the origin. For this reason we choose $\alpha(0) = 0$ and
$\partial_{r} \alpha|_{r=0} = 1$, since this corresponds to the $m$
extra dimensions taking the form

\begin{equation} ds^{2}_{[ m ]} = -dr^{2} - \alpha^{2}(r) \gamma_{ab}
dy^{a}dy^{b} = -dr^{2} - r^{2} d\Omega^{2}_{[m-1]},
\end{equation}

\noindent at $r=0$. In other words, the extra dimensions are simply
Minkowski space at the origin, and hence the metric is regular here. Integrating ($\ref{eq:b90}$) and using these boundary conditions, we obtain

\begin{equation}\label{eq:b9}  K_{A}{}^{B}|_{\epsilon} = {2M_{\rm b}^{m-1} \over
\Omega^{[m-1]}M^{n-2}}\left( \tilde{T}_{A}{}^{B} - \delta_{A}{}^{B}{\tilde{T}
\over n-2}\right) ,
\end{equation}

\noindent where we have defined $M_{\rm b} = \alpha(\epsilon)^{-1}$. This result has been derived previously in Refs.
\cite{n01}, \cite{n02} for codimension one and two branes, and an
explanation as to the relationship between these total derivatives
and the underlying symmetries of the background space is given in
Ref. \cite{rn00}. Note that if we consider the case
$m=1$, and take the limit $\epsilon \to 0$ we recover the five dimensional junction conditions.

We now consider ($\ref{eq:b7}$). Unlike ($\ref{eq:b6}$), there is no total derivative in
($\ref{eq:b7}$), due to the presence of the $\sqrt{G}
R^{(\gamma)}{}_{a}{}^{b} / \alpha^{2}$ term. To proceed, we
integrate ($\ref{eq:b7}$) over the region of the transverse space occupied by
the brane,

\begin{equation}\label{eq:b27} {1 \over 2}\Omega^{[m-1]} \int_{0}^{\epsilon}
(\sqrt{G}L_{a}{}^{b})' d r = {\Omega^{[m-1]} \over 2 M_{\rm
b}^{m-1}}\sqrt{g}|_{\epsilon} L_{a}{}^{b}|_{\epsilon} = {1 \over
M^{n-2}}\int d^{m-1}y dr \sqrt{G}\left(T_{a}{}^{b} -
\delta_{a}{}^{b}{T \over n-2}\right) - (m-2)\delta_{a}^{b}\int
d^{m-1}y dr \sqrt{G}{1 \over \alpha^{2}} .
\end{equation}

\noindent which is valid for $m > 2$. The problematic term is the
last one on the right hand side of ($\ref{eq:b27}$). Since we are
only integrating over the small transverse region occupied by the
brane (that is in the region $r<\epsilon $), we can find an
approximate solution to the above equation. To do so, we note that
we have chosen our boundary conditions at $r=0$ such that $\alpha(r)
= r$ for small $r$. In addition, we also impose the boundary
condition ${d f \over dr}|_{r=0} = 0$, which is also required for a
regular solution. We will assume that over the small brane region
$(0,\epsilon)$, that $\alpha \approx r$ and $f(r) \approx {\rm
const}$. In making this approximation we find

\begin{equation} \label{eq:nc1} L_{a}{}^{b}|_{r=\epsilon} \approx {2M_{\rm b}^{m-1}
\over \Omega^{[m-1]}M^{n-2}}\left( \tilde{T}_{a}{}^{b} -
\delta_{a}{}^{b}{\tilde{T} \over n-2} \right) - {2 \over
\epsilon}\delta_{a}{}^{b}  ,\end{equation}

\noindent which is valid for $m > 2$. For the cases $m=1,2$, the
last integral in ($\ref{eq:b27}$) is not defined; it diverges at
$r=0$. However, this term is absent when we consider
codimension one and two branes. We
will consider the specific examples of $m=1,2$ shortly.

We now have approximate expressions for $K_{A}{}^{B}$ and
$L_{a}{}^{b}$ at $r=\epsilon$, ($\ref{eq:b9}$) and ($\ref{eq:nc1}$). The next step is to consider the
$(r,r)$ Einstein equation,

\begin{equation}\label{eq:mm2} G_{r}{}^{r} = R_{r}{}^{r} - {1 \over 2}R
=    {1 \over 8}\left(K_{C}{}^{D}K_{D}{}^{C} +
L_{a}{}^{b}L_{a}{}^{b}\right) - {1 \over 8}\left( L^{2} +
K^{2}\right) - {1 \over 4}KL - {1 \over 2}\left({R^{(g)} \over f} +
{R^{(\gamma)} \over \alpha^{2}}\right) = {T_{r}{}^{r}|_{{\rm bulk}}
\over M^{n-2}}.
\end{equation}

\noindent To obtain an effective four dimensional field equation from ($\ref{eq:mm2}$), we should perform our regularization procedure and integrate it over the range $r=(0,\epsilon)$. However, for simplicity we will instead evaluate ($\ref{eq:mm2}$) at the surface $r = \epsilon$. We stress that this is only an approximation, and we should apply our regularization procedure to ($\ref{eq:mm2}$). However, given our particular profile function, we expect that evaluating ($\ref{eq:mm2}$) at $r=\epsilon$ is sufficient to obtain an approximate brane equation.
The reason why we use this approximation is that by virtue of the `junction conditions' ($\ref{eq:nc1}$) and
($\ref{eq:b9}$), we know $K_{A}{}^{B}$ and $L_{a}{}^{b}$ in terms of
the brane energy momentum tensor $\tilde{T}_{\mu}{}^{\nu}$ at
$r=\epsilon$. We consider the cases $m=1$, $m=2$ and $m>2$
separately. \medskip

\subsection{Codimension one}

For $m=1$ there are no $L_{a}{}^{b}$ components of the extrinsic
curvature, and in addition there is no $R^{(\gamma)}/ \alpha^{2}$
term in ($\ref{eq:mm2}$). Therefore using ($\ref{eq:b9}$) and
evaluating ($\ref{eq:mm2}$) at $r=\epsilon$, we find

\begin{equation}\label{eq:np197} - {1 \over 2}{R^{(g)} \over f(\epsilon)}  +
{A_{1}^{2} \over 8 (n-2)^{2}} \left[
\left((n-2)^{2}\tilde{T}_{A}{}^{B}\tilde{T}_{B}{}^{A} +
B_{1}(\tilde{T}_{A}{}^{A})^{2}\right) + \left(2B_{1}
\tilde{T}_{r}{}^{r} \tilde{T}_{A}{}^{A} +
C_{1}(\tilde{T}_{r}{}^{r})^{2}\right)\right] = -{\Lambda \over
M^{n-2}} ,\end{equation}

\noindent where the constants $A_{1}$, $B_{1}$ and  $C_{1}$ are
given by

 \begin{equation} A_{1}
= {1  \over  M^{n-2}}, \qquad \qquad \qquad B_{1}  = 1-p, \qquad
\qquad \qquad C_{1} = p-p^{2},
\end{equation}

\noindent We note that the brane energy momentum tensor now has a non-zero component $\tilde{T}_{r}^{r}$ in the transverse direction. This component is zero for a thin brane.

\subsection{Codimension two}

The case $m=2$
is unique; to see why we return to equation ($\ref{eq:b7}$), and
integrate over the region $r<\epsilon$. Evaluating this integral
yields two terms, one at $r=0$ and the other at $r=\epsilon$,

\begin{equation} \label{eq:bb1}{\Omega^{[m-1]} \over 2}
\int_{0}^{\epsilon} (\sqrt{G}L_{a}{}^{b})' dr = {\Omega^{[m-1]} \over
2}\left[ (\sqrt{G}L_{a}{}^{b})|_{r=\epsilon} -
(\sqrt{G}L_{a}{}^{b})|_{r=0}\right] .\end{equation}

\noindent When $m>2$ the term $(\sqrt{G}L_{a}{}^{b})|_{r=0}$ in
($\ref{eq:bb1}$) is zero by virtue of the boundary condition
$\alpha(0)=0$. However, for $m=2$, $(\sqrt{G}L_{a}{}^{b})|_{r=0}
\neq 0$ and we have to include a boundary term at the origin. Hence,
accounting for this extra term, equation ($\ref{eq:mm2}$), evaluated
at $r=\epsilon$, is

\begin{equation}\label{eq:np10} - {1 \over 2}{R^{(g)} \over
f(\epsilon)}  +  {A_{2}^{2} \over 8 (n-2)^{2}} \left[
\left((n-2)^{2}\tilde{T}_{A}{}^{B}\tilde{T}_{B}{}^{A} +
B_{2}(\tilde{T}_{A}{}^{A})^{2}\right) +
\left((n-2)^{2}\tilde{T}_{a}{}^{b}\tilde{T}_{b}{}^{a} +
B_{2}(\tilde{T}_{a}{}^{a})^{2}\right) \right]
\end{equation} \begin{equation*}+{A_{2}^{2} \over 8 (n-2)^{2}} \left[2B_{2}\left(  \tilde{T}_{a}{}^{a} +
\tilde{T}_{r}{}^{r}\right) \tilde{T}_{A}{}^{A} + 2B_{2}
\tilde{T}_{a}{}^{a}\tilde{T}_{r}{}^{r} +
C_{2}(\tilde{T}_{r}{}^{r})^{2}\right] \end{equation*} \begin{equation*} +{1 \over 2}A_{2} M_{\rm
b}^{m-1}{\sqrt{g}|_{r=0} \over
\sqrt{g}|_{r=\epsilon}}\tilde{T}_{a}{}^{a} + {1 \over 2}A_{2}
M_{b}^{m-1}{\sqrt{g}|_{r=0} \over \sqrt{g}|_{r=\epsilon}}
\tilde{T}_{r}{}^{r} = -{\Lambda \over M^{n-2}},\end{equation*}

\noindent where the constants $A_{2}$, $B_{2}$ and $C_{2}$ are given
by

 \begin{equation} A_{2}
= { M_{{\rm b}} \over \pi M^{n-2}}, \qquad \qquad \qquad B_{2}  =
-p, \qquad \qquad \qquad C_{2} = p-p^{2} -2p.
\end{equation}

\noindent We have confirmed that equations ($\ref{eq:np197}$),
($\ref{eq:np10}$) agree with the results of Refs. \cite{n01} and
\cite{n02}.

\subsection{Higher codimension}

Now, the term $R^{(\gamma)}/\alpha^{2}$ is not zero, and we find the
relation

\begin{equation}\label{eq:np100} - {R^{(g)} \over
2 f(\epsilon)}  +  {A_{m}^{2} \over 8 (n-2)^{2}}
\left[\left((n-2)^{2}\tilde{T}_{A}{}^{B}\tilde{T}_{B}{}^{A} +
B_{m}(\tilde{T}_{A}{}^{A})^{2}\right) +
\left((n-2)^{2}\tilde{T}_{a}{}^{b}\tilde{T}_{b}{}^{a} +
B_{m}(\tilde{T}_{a}{}^{a})^{2}\right)\right]
\end{equation} \begin{equation*}+ {A_{m}^{2} \over 8 (n-2)^{2}}\left[ 2B_{m}\left(  \tilde{T}_{a}{}^{a} +
\tilde{T}_{r}{}^{r}\right) \tilde{T}_{A}{}^{A} + 2B_{m}
\tilde{T}_{a}{}^{a}\tilde{T}_{r}{}^{r} +
C_{m}(\tilde{T}_{r}{}^{r})^{2} \right] \end{equation*} \begin{equation*}- {A_{m} \over
2\epsilon}\tilde{T}_{a}{}^{a} - {A_{m}(m-1) \over
2\epsilon}\tilde{T}_{r}{}^{r} - {(m-1)(m-2) \over \epsilon^{2}}=
-{\Lambda \over M^{n-2}}.\end{equation*}

\noindent where $A_{m}$, $B_{m}$ and $C_{m}$ are given by

 \begin{equation} A_{m}
= {2 M_{{\rm b}}^{m-1} \over \Omega^{[m-1]} M^{n-2}}, \qquad \qquad
\qquad B_{m}  = 2-m-p,  \qquad \qquad \qquad C_{m} =
p-p^{2}+(m-1)-(m-1)^{2} -2p(m-1), \end{equation}

\noindent for $m>2$. The equation ($\ref{eq:np100}$) gives the brane
Ricci scalar $R^{(g)}$ in terms of the bulk cosmological constant
$\Lambda$ and the 'four dimensional' energy momentum tensor
$\tilde{T}_{\mu\nu}$. We will return to equations ($\ref{eq:np197}$),
($\ref{eq:np10}$) and ($\ref{eq:np100}$) shortly.

\medskip

Next, we consider the $(A,a)$, $(A,r)$ components of $R_{\mu\nu}$
evaluated at $r=\epsilon$, which are simply given by $R_{A a} = 0$,
and $ R_{A r} = 0$. For our particular metric, $R_{A r}=0$ tells us
that $G_{A r} = 0$, which implies $T_{A r}|_{{\rm bulk}} = 0$. As
this component of the energy momentum tensor describes the flow of
energy from the brane to the bulk, it appears that for braneworld
models with a metric given by ($\ref{eq:1}$) we will obtain no loss
of energy into the bulk. However this conclusion is not necessarily
true if the brane thickness is variable, as was discussed in Ref.
\cite{n01}.

\section{Special cases of static braneworlds}

The equations ($\ref{eq:np197}$),
($\ref{eq:np10}$) and ($\ref{eq:np100}$) that were derived in the previous section can act as generalized fine tuning relations for branes of codimension $m=1$, $m=2$ and $m>2$
respectively. To see this, we set $R^{(g)} = 0$. Taking the $m=1$
case as an example, we therefore see that the fine tuning relation

\begin{equation}\label{eq:ft1}
 \left[
\left((n-2)^{2}\tilde{T}_{A}{}^{B}\tilde{T}_{B}{}^{A} +
B_{1}(\tilde{T}_{A}{}^{A})^{2}\right) + \left(2B_{1} \tilde{T}_{r}{}^{r}
\tilde{T}_{A}{}^{A} + C_{1}(\tilde{T}_{r}{}^{r})^{2}\right)\right] = -{8
(n-2)^{2} \over A_{1}^{2}  M^{n-2}} \Lambda
\end{equation}

\noindent must hold to ensure the brane is flat. If we take the
background space to be five dimensional, and consider a pure tension
brane, then when we take the $\epsilon \to 0$ limit, we find
$\tilde{T}_{r}{}^{r} \to 0$ and $\tilde{T}_{A}{}^{B} \to -{1 \over
2}\delta_{A}{}^{B} \lambda$, where $\lambda$ is the brane tension.
Using these results in ($\ref{eq:ft1}$), we find the fine tuning
relation

\begin{equation}\label{eq:ft2} \Lambda = {\lambda^{2} \over 24
M^{3}},
\end{equation}

\noindent which is equivalent to the standard Randall Sundrum fine
tuning condition. There is a sign discrepancy between this fine tuning and the RS model, but this is because we are considering the metric,

\begin{equation} ds^{2}_{[5]} = f(r) g_{\mu\nu}dx^{\mu}dx^{\nu} -
dr^{2} .\end{equation}

\noindent which is five dimensional de Sitter space, not AdS as in
the RS model. If we consider an AdS space, we set $\Lambda \to -\Lambda$ and the correct
fine tuning is recovered.

Whilst ($\ref{eq:ft2}$) is valid for a thin brane, for a thick brane
we have the fine tuning relation ($\ref{eq:ft1}$), which we expect
will depend on the brane profile. For a pure tension brane, we can
write $\hat{T}_{A}^{B} = -{1 \over 2}\lambda \delta_{A}^{B}
D_{\epsilon}(y)$, where $D_{\epsilon}(y)$ is the brane profile. If
we normalize the brane profile appropriately such that

\begin{equation} \int_{-\epsilon}^{\epsilon} \sqrt{G}
D_{\epsilon}(y) dy = 1 ,\end{equation}

\noindent then we can write the effective four dimensional energy
momentum tensor as $\tilde{T}_{A}^{B} = -{1 \over 2} \lambda
\delta_{A}^{B}$, as before. It follows that the fine tuning relation
($\ref{eq:ft1}$) has the form

\begin{equation} \Lambda =  {\lambda^{2} \over 24M^{3}} - {
\lambda \tilde{T}_{r}{}^{r} \over 6M^{3}} +
{(\tilde{T}_{r}{}^{r})^{2} \over 6M^{3}}. \end{equation}

\noindent We see for a thick brane the fine tuning relation is modified by the
 $\tilde{T}_{r}{}^{r}$ components of the brane
energy momentum tensor.

We can continue in this manner and consider the codimension two
case. We first consider a thin, pure tension codimension two brane for which
we have $\tilde{T}_{a}{}^{b} = \tilde{T}_{r}{}^{r} =
0$. For pure tension branes, we can use the property
$\tilde{T}_{A}{}^{B} = \delta_{A}{}^{B} \tilde{T}_{C}{}^{C}/4$. If
we use this relation, and set $m=2$, $p=4$ and $n=6$ in
($\ref{eq:np10}$), we find that if we take $R^{(g)} = 0$, then all of the terms on the
left hand side are zero. We are left simply with $\Lambda = 0$ at
the position of the brane. If we consider instead a more general
bulk energy momentum tensor $T_{r}{}^{r}|^{\rm bulk}$, then this
result becomes $T_{r}{}^{r}|^{\rm bulk} = 0$. In \cite{sc1}, a six
dimensional model was considered, and it was found that the bulk
energy momentum tensor had to be tuned in such a way that
$T_{r}{}^{r}|^{\rm bulk}=0$ in order to have a flat brane. Hence our
result is in agreement with \cite{sc1}, and reflects that fact that
to obtain a flat pure tension codimension two brane, the brane tension
 does not have to be fine tuned to the bulk
energy momentum tensor.

Next, we consider a thick brane, so we cannot assume
$\tilde{T}_{a}{}^{b}=\tilde{T}_{r}{}^{r} = 0$. We find that
($\ref{eq:np10}$) now reads

\begin{equation}\label{eq:ff10} {A_{2}^{2} \over 8 (n-2)^{2}} \left[
\left((n-2)^{2}\tilde{T}_{a}{}^{b}\tilde{T}_{b}{}^{a} +
B_{2}(\tilde{T}_{a}{}^{a})^{2}\right)+ 2B_{2}\left(
\tilde{T}_{a}{}^{a} + \tilde{T}_{r}{}^{r}\right) \tilde{T}_{A}{}^{A}
+ 2B_{2} \tilde{T}_{a}{}^{a}\tilde{T}_{r}{}^{r} +
C_{2}(\tilde{T}_{r}{}^{r})^{2} \right] +
\end{equation} \begin{equation*}  +{1 \over 2}A_{2} M_{\rm
b}^{m-1}{\sqrt{g}|_{r=0} \over
\sqrt{g}|_{r=\epsilon}}\tilde{T}_{a}{}^{a} + {(m-1) \over 2}A_{2}
M_{{\rm b}}^{m-1}{\sqrt{g}|_{r=0} \over \sqrt{g}|_{r=\epsilon}}
\tilde{T}_{r}{}^{r} = -{\Lambda \over M^{n-2}}.\end{equation*}

\noindent Now if we assume that the $\tilde{T}_{a}{}^{b}$ and
$\tilde{T}_{r}{}^{r}$ components of the brane energy momentum tensor
are small, then we find the following approximate fine tuning
relation that must be satisfied in order for a thick codimension two
brane to be flat,

\begin{equation} \label{eq:ff11} -{M_{\rm b}^{2} \over 16
\pi^{2}M^{8}}\left(  \tilde{T}_{a}{}^{a} +
\tilde{T}_{r}{}^{r}\right) \tilde{T}_{A}{}^{A} + {M_{\rm b}^{2}
\over 2\pi M^{4}} {\sqrt{g}|_{r=0} \over \sqrt{g}|_{r=\epsilon}}
\left(\tilde{T}_{a}^{a} + \tilde{T}_{r}^{r}\right) = -{\Lambda \over
M^{4}}.\end{equation}

\noindent where we have taken $p=4$. We see that to obtain a flat
brane in this case, we must tune $\tilde{T}_{A}{}^{A}$ to the bulk
energy momentum tensor (unless we have a model in which
$\tilde{T}_{a}{}^{a} = - \tilde{T}_{r}{}^{r}$, in which case we have
$\Lambda \approx 0$, or more generally $T_{r}{}^{r}|^{\rm bulk} =
0$, as in the thin brane case.)

\section{\label{sec:1}Cosmological Solution}

Having considered metrics of the form ($\ref{eq:1}$) in the previous
section, we now perform an analogous calculation for the time-dependent
metric ($\ref{eq:197}$). We initially incorporate a time dependence
in the extra dimensions by writing $\alpha(r,t)$ as a function of
time. However, to perform the same calculation as above, we find
that we must remove this time dependence. This amounts to assuming
that the extra dimensions are static.

Calculating the Ricci tensor $R_{\mu}{}^{\nu}$ for the metric
($\ref{eq:197}$) as we did in the previous section gives

\begin{equation}\label{eq:c1} R_{i}{}^{j} =
{1 \over 2}{\left(N A^{p-1} \alpha^{m-1} K_{i}{}^{j}\right)' \over N
A^{p-1}\alpha^{m-1}} +  {R^{(g)}{}_{i}{}^{j} \over A^{2}} - {(m-1)
\over \alpha}\nabla^{j}\nabla_{i}\alpha = {1 \over
M^{n-2}}\left(T_{i}{}^{j} - \delta_{i}{}^{j}{T \over
n-2}\right),\end{equation}

\begin{equation}\label{eq:c2}R_{t}{}^{t} =
{1 \over 2}{\left(N A^{p-1} \alpha^{m-1} K_{t}{}^{t}\right)' \over N
A^{p-1}\alpha^{m-1}} +  {R^{(g)}{}_{t}{}^{t} \over A^{2}}  - {(m-1)
\over \alpha}\nabla^{t}\nabla_{t}\alpha = {1 \over
M^{n-2}}\left(T_{t}{}^{t} - \delta_{t}{}^{t}{T \over
n-2}\right),\end{equation}

\begin{equation}\label{eq:c3} R_{a}{}^{b} = {1 \over
2}{\left(N A^{p-1}\alpha^{m-1} L_{a}{}^{b}\right)' \over N
A^{p-1}\alpha^{m-1}} - {(m-2) \over \alpha^{2}}\delta_{a}{}^{b}
\nabla_{\mu} \alpha \nabla^{\mu} \alpha - {\delta_{a}{}^{b} \over
\alpha} \nabla^{\mu}\nabla_{\mu}\alpha + {(m-2) \over
\alpha^{2}}\delta_{a}{}^{b}
\end{equation} \begin{equation*}= {1 \over M^{n-2}}\left(T_{a}{}^{b} -
\delta_{a}{}^{b}{T \over n-2}\right),
\end{equation*}

\begin{equation} R_{r}{}^{r} = {1 \over 2}\left( K' + L'\right)
+ {1 \over 4} \left(K_{A}{}^{B}K_{B}{}^{A} +
L_{a}{}^{b}L_{b}{}^{a}\right) .
\end{equation}

\noindent where the extrinsic curvatures are defined in
($\ref{eq:si1}$). Primes denote derivatives with respect to the
radial coordinate $r$, and dots derivatives with respect to time. We
have defined the Ricci tensors of the metrics $g_{ij}$ and
$\gamma_{ab}$ as $R^{(g)}{}_{i}{}^{j}$ and
$R^{(\gamma)}{}_{a}{}^{b}$ respectively, and the covariant
derivatives $\nabla_{\mu}$ preserve the metric $g_{\mu\nu}$.

The final equation that we will need is

\begin{equation} \label{eq:b15} G_{A r} = R_{A r} = - {
(m-1)\partial_{A} \alpha' \over \alpha} + {(m-1) \partial_{B} \alpha
\over 2\alpha} K^{B}{}_{A} + {1 \over 2} \nabla^{B}(K_{AB} - g_{AB}K
)= {T_{A r}^{{\rm bulk}} \over M^{n-2}} .\end{equation}

\noindent which will give us a
conservation equation for the brane energy momentum tensor
$\tilde{T}_{\mu\nu}$.

We will now make a number of assumptions in order to obtain an
equation for $A(t,r)$ at the surface of the brane. We begin as
before, by assuming that brane tangential derivatives are small, in
particular we will neglect the terms $\nabla^{j}\nabla_{i}\alpha$
and $\nabla_{t}\nabla^{t}\alpha$ in equations ($\ref{eq:c1}$),
($\ref{eq:c2}$) and ($\ref{eq:c3}$). With this assumption we have
 removed the time dependence of the extra dimensions, and
so our solution will be valid only when the transverse dimensions
are static or weakly time dependent.

We proceed as in the previous section. We neglect derivatives tangential to the brane,
and integrate ($\ref{eq:c1}$-$\ref{eq:c3}$) over the brane thickness $0<r<\epsilon$, using the approximations $N(t,r=0) \approx N(t,r=\epsilon)=1$ and $\alpha(t,r<\epsilon)
\approx r$. We find

\begin{equation}\label{eq:b17} K_{i}{}^{j}|_{r=\epsilon} = {2 M_{\rm b}^{m-1}
\over \Omega^{[m-1]}M^{n-2}}\left(\tilde{T}_{i}{}^{j} -
\delta_{i}{}^{j}{\tilde{T} \over n-2}\right) ,\end{equation}

\begin{equation}\label{eq:b18} K_{t}{}^{t}|_{r=\epsilon} = {2 M_{\rm b}^{m-1}
\over \Omega^{[m-1]}M^{n-2}}\left(\tilde{T}_{t}{}^{t} -
\delta_{t}{}^{t}{\tilde{T} \over n-2}\right) ,\end{equation}

\begin{equation}\label{eq:b19} L_{a}{}^{b}|_{r=\epsilon} = {2 M_{\rm b}^{m-1}
\over \Omega^{[m-1]}M^{n-2}}\left(\tilde{T}_{a}{}^{b} -
\delta_{a}{}^{b}{\tilde{T} \over n-2}\right) -{2 \over
\epsilon}\delta_{a}{}^{b},\end{equation}

\noindent which are valid for $m \ge 3$. As in the previous section, for
codimension two objects we obtain an additional term in
($\ref{eq:b19}$), a boundary term at $r=0$ of the form
$\left(\sqrt{G} L_{a}{}^{b}\right)|_{r=0}$. We will check throughout
though that our results can give the results of \cite{n01} for a
codimension two brane.

The next step is to evaluate the $(r,r)$ Einstein equation at
$r=\epsilon$,

\begin{equation}\label{eq:b20} G_{r}{}^{r}|_{r=\epsilon} = -{1 \over 2}R^{(g)}-{1
\over 2}{(m-1)(m-2) \over \alpha^{2}} + {1 \over 8}\left(
K_{C}{}^{D}K_{D}{}^{C} - K^{2}\right) + {1 \over 8}\left(
L_{a}{}^{b}L_{b}{}^{a} - L^{2} \right) - {1 \over 4}L K = -{\Lambda
\over M^{n-2}} .\end{equation}

\noindent We assume that $A(r=0) \approx A(r=\epsilon) = a(t)$ is
the brane scale factor. We also set the energy momentum tensor on
the brane $\tilde{T}_{\mu}{}^{\nu}$ as

\begin{equation}\label{eq:1000} \tilde{T}_{\mu}{}^{\nu} = {\rm
diag}\left(\rho,-p_{\rm br},-p_{\rm br},-p_{\rm br},-p_{\rm r},-p_{\rm
bk},..,-p_{\rm bk}\right) ,\end{equation}

\noindent where $\rho$ is the energy density, $p_{\rm br}$ the
brane pressure components, and $p_{\rm r}$ and $p_{\rm bk}$ the
extra dimensional pressure components. There are $(m-1)$ $p_{\rm
bk}$ terms in ($\ref{eq:1000}$). Using ($\ref{eq:1000}$) in ($\ref{eq:b17}$-$\ref{eq:b19}$), the matching conditions become

\begin{equation}\label{eq:b21} K_{i}{}^{j}|_{r=\epsilon} = -{2 M_{\rm b}^{m-1}
\over \Omega^{[m-1]}M^{n-2}(n-2)}\delta_{i}{}^{j}\left[
(n-1-p)p_{\rm br} +\rho - p_{\rm r} - (m-1)p_{\rm bk} \right] ,
\end{equation}

\begin{equation}\label{eq:b22} K_{t}{}^{t}|_{r=\epsilon} = {2 M_{\rm b}^{m-1}
\over \Omega^{[m-1]}M^{n-2}(n-2)}\left[ (n-3)\rho + (p-1)p_{\rm br}
+ p_{\rm r} + (m-1)p_{\rm bk}  \right] , \end{equation}

\begin{equation}\label{eq:b23} L_{a}{}^{b}|_{r=\epsilon} = -{2 M_{\rm b}^{m-1}
\over \Omega^{[m-1]}M^{n-2}(n-2)}\delta_{a}{}^{b}\left[
(n-1-m)p_{\rm bk} + \rho - (p-1)p_{\rm br} -p_{\rm r}  \right] -{2
\over \epsilon}\delta_{a}{}^{b} ,\end{equation}

\noindent and the $G_{r}^{r}$ Einstein equation is,

\begin{equation} \label{eq:b25} G_{r}{}^{r}|_{r=\epsilon} =
3\left({\ddot{a}\over a}+ \left({\dot{a}
\over a}\right)^{2}\right)+ {A_{m}^{2} \over 8 (n-2)^{2}}
\left((n-2)^{2}\tilde{T}_{A}{}^{B}\tilde{T}_{B}{}^{A} +
B_{m}(\tilde{T}_{A}{}^{A})^{2} +
(n-2)^{2}\tilde{T}_{a}{}^{b}\tilde{T}_{b}{}^{a} +
B_{m}(\tilde{T}_{a}{}^{a})^{2}\right)
\end{equation} \begin{equation*}+ {A_{m}^{2} \over 8 (n-2)^{2}}\left(2B_{m}\left(  \tilde{T}_{a}{}^{a} +
\tilde{T}_{r}{}^{r}\right) \tilde{T}_{A}{}^{A} + 2B_{m}
\tilde{T}_{a}{}^{a}\tilde{T}_{r}{}^{r} +
C_{m}(\tilde{T}_{r}{}^{r})^{2}\right) - {A_{m} \over 2
\epsilon}\tilde{T}_{a}{}^{a} - {(m-1)A_{m} \over 2\epsilon}
\tilde{T}^{r}{}_{r}- { (m-1)(m-2)\over \epsilon^{2}} = -{\Lambda
\over M^{n-2}}.\end{equation*}

Now, using the brane energy momentum tensor ($\ref{eq:1000}$), we obtain the following equation for the brane scale factor

\begin{equation}\label{eq:b30} 3\left({\ddot{a}\over a}+ \left({\dot{a}
\over a}\right)^{2}\right) = -{\Lambda \over M^{n-2}} - {A_{m}^{2}
\over 8(m+2)}\left[(1+m)(\rho + p_{{\rm br}} )^{2} + (2m-4) p_{{\rm
br}} ( p_{{\rm br}} - \rho) +3(m-1) p_{{\rm bk}}^{2} \right]+
\end{equation} \begin{equation*}-{A_{m}^{2} \over
8(m+2)} \left[ (2(m-1) p_{{\rm bk}} +2 p_{{\rm r}} )(\rho - 3
p_{{\rm br}} ) - 2(m-1) p_{{\rm bk}} p_{{\rm r}} - (m+3) p_{{\rm
r}}^{2} \right] \end{equation*}
\begin{equation*}+ {(m-1)(m-2) \over \epsilon^{2}} -
{(m-1)A_{m} \over 2\epsilon}(p_{\rm r}+p_{\rm bk}) .
\end{equation*}

This equation describes the evolution of the scale factor $a(t)$ at
the surface $r = \epsilon$ of a four dimensional, codimension $m$
brane, subject to the approximations and assumptions that we have
made thus far. Equation ($\ref{eq:b30}$) is valid for $m>2$; as we
have discussed in the previous section the codimension one and two
cases are unique and should be considered separately. Before
continuing with the general case, we briefly discuss the previously
studied $m=1,2$ examples.

For the codimension one case, there is no $L_{a}{}^{b}$ field
equation, and ($\ref{eq:b30}$) reads

\begin{equation}\label{eq:bb30} 3\left({\ddot{a}\over a}+ \left({\dot{a}
\over a}\right)^{2}\right) = -{\Lambda \over M^{n-2}} - {1 \over
12M^{6}}\left[(\rho + p_{{\rm br}} )^{2} - p_{{\rm br}} ( p_{{\rm
br}} - \rho) +   p_{{\rm r}} (\rho - 3 p_{{\rm br}} )  - 2 p_{{\rm
r}}^{2} \right],\end{equation}

\noindent which agrees with the results of Ref. \cite{n02}, where a
thick codimension one brane was considered. Next, we consider the
codimension two case. Once again, a slight complication arises from
a boundary term at $r=0$. We write the modified $(a,b)$
field equation as

\begin{equation}\label{eq:md23} L_{a}{}^{b}|_{r=\epsilon} = -{ M_{{\rm b}}
\over 4\pi M^{4}}\delta_{a}{}^{b}\left[ 3p_{\rm bk} + \rho - 3p_{\rm
br} -p_{\rm r}  \right] + {\sqrt{g}|_{r=0} \over
\sqrt{g}|_{r=\epsilon}}M_{{\rm b}},\end{equation}

\noindent and hence for $m=2$ the equation ($\ref{eq:b20}$) becomes

\begin{equation}\label{eq:bbb30} 3\left({\ddot{a}\over a}+ \left({\dot{a}
\over a}\right)^{2}\right) = -{\Lambda \over M^{4}} - {M_{{\rm
b}}^{2} \over 32\pi^{2}M^{8}}\left[3(\rho + p_{{\rm br}} )^{2}
 +3
p_{{\rm bk}}^{2} + 2(p_{{\rm bk}} +p_{{\rm r}})(\rho - 3 p_{{\rm
br}} ) - 2 p_{{\rm bk}} p_{{\rm r}} - 5 p_{{\rm r}}^{2} \right]
 \end{equation} \begin{equation*}+
{M_{{\rm b}}^{2} \over 2\pi M^{4}}(p_{\rm r}+p_{\rm
bk}){\sqrt{g}|_{r=0} \over \sqrt{g}|_{r=\epsilon}}
\end{equation*}

\noindent Once again, this result agrees with the results in Ref.
\cite{n01}.

Having checked that our equation ($\ref{eq:b30}$) agrees with the
well studied cases $m=1,2$, we can now consider the general case $m>2$. The brane energy momentum tensor contains the usual four
dimensional energy density and pressure terms $\rho$ and $p_{{\rm
br}}$, and in addition non-zero pressure terms $p_{r}$ and $p_{{\rm
bk}}$ in the transverse directions. If we assume that $p_{{\rm r}}$
and $p_{{\rm bk}}$ are constant across the brane, we can write
($\ref{eq:b30}$) as

\begin{equation} \label{eq:b31} 3\left({\ddot{a}\over a}+ \left({\dot{a}
\over
a}\right)^{2}\right) = \omega_{1} + \omega_{2}(\rho + p_{{\rm br}}
)^{2} + \omega_{3}p_{{\rm br}} ( p_{{\rm br}} - \rho) +
\omega_{4}(\rho - 3 p_{{\rm br}} )
\end{equation}

\noindent where the constants $\omega_{1,2,3,4}$ are given by

 \begin{align} \omega_{1} &= -{\Lambda \over
M^{n-2}} - {A_{m}^{2} \over 8(m+2)} \left[ 3(m-1)p_{{\rm bk}}^{2} -
2(m-1)p_{{\rm bk}}p_{{\rm r}} - (m+3)p_{{\rm r}}^{2} \right] \\
&+ {(m-1)(m-2) \over \epsilon^{2}} -
{(m-1)A_{m} \over 2\epsilon}(p_{r}+p_{bk}), \\
\omega_{2} &= -{A_{m}^{2}(1+m) \over 8 (m+2)} , \\
 \omega_{3} &= -{ A_{m}^{2}(m-2) \over 4 (m+2)}, \\
 \omega_{4} &= -{  A_{m}^{2}  \over 4
(m+2)}[(m-1)p_{{\rm bk}} + p_{{\rm r}}] .\end{align}

\noindent From ($\ref{eq:b30}$), we see that in addition to the standard $(\rho
- 3p_{{\rm br}})$ term, we also have the quadratic terms $(\rho
+ p_{{\rm br}})^{2}$ and $p_{{\rm br}}(p_{{\rm br}} -\rho)$. If we
take $\rho = T + \rho_{m}$ and $p_{{\rm br}} = -T + w \rho_{m}$,
where $\rho_{m}$ is a small energy component with arbitrary equation
of state $p_{m}=w \rho_{m}$ and $T$ a constant brane tension, we can expand ($\ref{eq:b31}$) in
powers of $\rho_{m}$. Doing so we obtain

\begin{equation} \label{eq:b32} 3\left({\ddot{a}\over a}+ \left({\dot{a}
\over a}\right)^{2}\right) = (\omega_{1} +2T^{2}\omega_{3} +
4\omega_{4}T) + (\omega_{4} + T\omega_{3})(1 - 3w )\rho_{m} +
O(\rho_{m}^{2}),
\end{equation}

\noindent which is the standard cosmological equation when
considering late time cosmology (that is when $\rho_{m}$ is small).
The cosmological constant problem in this particular model is why
the constant $(\omega_{1} +2T^{2}\omega_{3} + 4\omega_{4}T)$ is
either zero or very small. There is no reason to expect these terms
to cancel one another, suggesting these models are not free from
fine tuning. This is not a surprising result, since we had no reason
to expect that self tuning behaviour exists in these models.

Thus we have found that late-time cosmology with quadratic corrections is a generic feature of certain braneworld models, regardless of
the codimension, generalizing the arguments made in \cite{n01} and
\cite{n02} for codimension one and two branes. We note that since we
have assumed that $p_{r}$ and $p_{\rm bk}$ are constants, then we must necessarily get an expression of the form

\begin{equation} \label{eq:ex1} 3\left({\ddot{a} \over a}+\left({\dot{a} \over a}\right)^{2}\right) =
\kappa_{1}(p_{r},p_{\rm bk}) + \kappa_{2}(p_{r},p_{\rm bk})(\rho -
3p_{\rm br}) + \kappa_{3}(\rho^{2},p_{br}^{2},\rho p_{br}) +{\Lambda
\over M^{n-2}}
\end{equation}

\noindent where $\kappa_{1},\kappa_{2}$ are functions of $p_{r}$ and
$p_{\rm bk}$ only, and hence are constants, and $\kappa_{3}$ is a
term quadratic in the variables $\rho, p_{\rm br}$. It was remarked
in \cite{n01} that if we obtained a cosmological equation like

\begin{equation} \label{eq:ex2} 3\left({\ddot{a} \over a}+\left({\dot{a} \over a}\right)^{2}\right) =
F(\rho+p_{\rm br}) + G(p_{r},p_{\rm bk},\rho,p_{\rm br}) - {\Lambda
\over M^{n-2}}
\end{equation}

\noindent where $F$ and $G$ are functions of the brane energy
density, then if $F(\rho+p_{\rm br})$ was a function linear in
$(\rho + p_{\rm br})$ then we would have a potential mechanism for
self tuning. In our setup, the function $F(\rho +p_{\rm br})$ will
always be quadratic in $(\rho+p_{\rm br})$.

The above arguments apply when the bulk
energy momentum tensor $T_{r}^{r}|^{\rm bulk}$ is not a function of
$\rho, p_{\rm br}$ and $p_{r}, p_{\rm bk}$ are constants. If we instead assume that $p_{r}$ and
$p_{\rm bk}$ are related to $\rho$ by the equations of state $p_{r}
= w_{r}\rho$ and $p_{\rm bk} = w_{\rm bk}\rho$, then we would obtain
 a cosmological model that differs from ($\ref{eq:b32}$) \cite{n01}.

Next, we derive a conservation equation for the
brane energy momentum tensor $\tilde{T}_{\mu}^{\nu}$. We do so from
equation ($\ref{eq:b15}$), evaluated at $r=\epsilon$. Using our
matching conditions at the surface of the $n$-brane, we find that
($\ref{eq:b15}$) reads

\begin{equation}\label{eq:sww1} {A_{m} \over 2}\nabla^{B}\tilde{T}_{AB} - {A_{m} \over 2(n-2)}  (2-m-p)\nabla_{A}\tilde{T}_{r}^{r}
+{(m-1)A_{m} \partial_{B}M_{b} \over 2M_{b}} \left(\tilde{T}_{A}^{B}
- \delta_{A}^{B} {\tilde{T} \over n-2}\right) = {T_{A r}^{\rm
bulk}|_{r=\epsilon} \over M^{n-2}} .\end{equation}

\noindent We have assumed throughout this paper that the time
dependence of the extra dimensions can be neglected to our level of
approximation. This means that the $\partial_{A} M_{b}$ term in
($\ref{eq:sww1}$) can be neglected. In addition, we assumed that the
$\tilde{T}_{r}^{r}$ component of the brane energy momentum tensor
was approximately constant, which means that we can write
($\ref{eq:sww1}$) as

\begin{equation}\label{eq:sww2} {A_{m} \over 2}( \dot{\rho} + 3{\dot{a}
\over a}(\rho + p_{{\rm br}}))
 \approx {T_{t r}^{\rm bulk}|_{r=\epsilon} \over M^{n-2}},
\end{equation}

\noindent and hence, if we have $T_{t r}^{\rm bulk}|_{r=\epsilon} =
0$ then we obtain the standard four dimensional conservation
equation

\begin{equation}\label{eq:sww3}  \dot{\rho} + 3{\dot{a}
\over a}(\rho + p_{{\rm br}}) =0 ,\end{equation}

\noindent However, this is only approximate, subject to the
assumption that $\partial_{t} M_{b} \approx 0$. We also note that if
$T_{t r}^{\rm bulk}|_{r=\epsilon} \neq 0$, then the brane energy
momentum tensor is not strictly conserved, and we can obtain a flow
of energy into the bulk. However, if we assume that $T_{t r}^{\rm
bulk}|_{r=\epsilon} = 0$, then from ($\ref{eq:sww3}$) we can write

\begin{equation}\label{eq:swf1} \rho = \rho_{c} a^{-3(1+w)} ,\end{equation}

\noindent where $\rho_{c}$ is a constant, and $w = p_{{\rm br}} /
\rho$. If we use ($\ref{eq:swf1}$) in ($\ref{eq:b31}$), then by
multiplying the equation by $\dot{a}a^{3}$ we can write the left
hand side as a total derivative,

\begin{equation} {3 \over 2}{d \over dt}
\left(\left(\dot{a}a\right)^{2}\right) =
\omega_{1}\dot{a}a^{3} + \left(\omega_{2}(1+w)^{2}+\omega_{3}
w(w-1)\right)\rho_{c}^{2} \dot{a} a^{-6(1+w)+3} +
\omega_{4}(1-3w)\rho_{c}\dot{a} a^{-3(1+w)+3} .\end{equation}

\noindent This equation can be integrated to give a Friedmann type
equation for the brane scale factor,

\begin{equation} \label{eq:fr1} H^{2} = \left({ \dot{a} \over
a}\right)^{2} = {\omega_{1} \over 6} + {2\omega_{4}  \over 3} \rho +
{2\left[\omega_{2}(1+w)^{2}+\omega_{3} w(w-1)\right] \over
3[4-6(1+w)]}\rho^{2} + {C \over a^{4}},
\end{equation}

\noindent where $C$ is an integration constant. The first two terms
on the right hand side of ($\ref{eq:fr1}$) are what we would expect
for standard four dimensional cosmology. The third term is quadratic
in the brane energy momentum tensor, and the fourth term is a `dark
radiation energy' component. It is curious that we find a dark
radiation contribution from the bulk; we might expect that in higher
codimension additional terms would be present in the brane field
equations. We will return to this point in the appendix.

\section{\label{sec:100} Discussion}

To review our progress; we have written the full
$n$-dimensional Ricci tensor and scalar in terms of $R^{(g)}$,
$R^{(\gamma)}$ and the extrinsic curvatures $K_{A}{}^{B}$ and
$L_{a}{}^{b}$ (see ($\ref{eq:b6}$),($\ref{eq:b7}$) and
($\ref{eq:b8}$)). We then decomposed the total energy momentum
tensor into a brane component, strictly localized in the region
$r<\epsilon$, and a bulk cosmological constant. Since the brane energy momentum tensor is assumed to localized in the region $r<\epsilon$, we are considering a local defect.

Our work is a particular example of the method outlined in Ref.
\cite{c1}, where brane quantities are defined as the projection of
higher dimensional terms onto a four dimensional subspace. Doing so
for the subspace $r=\epsilon$, we have found the approximate
cosmological equation ($\ref{eq:b31}$). In our setup, we see that
the four dimensional Planck mass $M_{{\rm pl}}^{2}$ is given by
the coefficient of the $\rho$ term in ($\ref{eq:b32}$), that is

\begin{equation} \label{eq:mp32} M_{{\rm pl}}^{2} = {1
\over 2\left(\omega_{4} + T \omega_{3}\right)}.
\end{equation}

\noindent We note that our definition of the Planck mass differs
from the effective action approach considered in (amongst others) Ref.
\cite{gh00}, where it is rather defined as

\begin{equation}\label{eq:mp33} M_{\rm pl}^{2} = M^{m+2} \int \sqrt{G} d^{m}y ,\end{equation}

\noindent in other words an integral over the $m$ codimensional space, which for the metric considered in this paper is given by

\begin{equation} M_{\rm pl}^{2} = M^{m+2} \Omega^{[m-1]} \int \alpha^{m-1} dr . \end{equation}

\noindent  The difference between the definitions ($\ref{eq:mp32}$) and ($\ref{eq:mp33}$) reflects the two different philosophies adopted in the literature towards defining four dimensional quantities. In the effective action approach, four-dimensional quantities are obtained by integrating out the extra dimensions in the full $n$-dimensional action. However, in this paper we use a tensorial approach, and project $n$-dimensional tensors tangentially to the brane. For a discussion of the two approaches, see Ref. \cite{bt01}.

As discussed in Refs. \cite{gh00}, \cite{ben1}, the codimension $m \geq 3$
case is different to the more commonly studied $m=1,2$ cases in the
literature. To see the difference, we consider the $n$-dimensional
Ricci scalar for our static metric ($\ref{eq:1}$),

\begin{equation}\label{eq:ri1} R = K' + L' +{1 \over 4}\left(K_{A}{}^{B}K_{B}{}^{A} +
L_{a}{}^{b}L_{a}{}^{b}\right) + {1 \over 4} \left( K^{2} +
L^{2}\right) + {1 \over 2} KL + {R^{(g)} \over f} + {(m-1)(m-2)
\over \alpha^{2}}.
\end{equation}

\noindent We see that the last term in ($\ref{eq:ri1}$) is present
only when $m>2$, and represents the curvature of the bulk. We note
that this is singular if $\alpha=0$ anywhere in the transverse
space. Thus to avoid singularities in the curvature invariants, we
must impose $\alpha \neq 0$ for all $r>0$. Note that we have imposed $\alpha = 0$ as a boundary condition at $r=0$, however this is simply a coordinate singularity.

As has been discussed in Ref. \cite{gh00}, the last term in ($\ref{eq:ri1}$) is a problem when we consider localizing gravity on the brane. To see why, we consider the Einstein equations ($\ref{eq:b6}$),
($\ref{eq:b7}$) and ($\ref{eq:b8}$) for our metric ($\ref{eq:1}$).
In the asymptotic limit $r \to \infty$, we require a solution to the
Einstein equations that is Anti-de-Sitter, that is we require $R \sim
{\rm const}$. To obtain a solution of this form all terms on the
right hand side of ($\ref{eq:ri1}$) must either asymptote to zero
or a constant as $r \to \infty$. We are considering flat
branes, so $R^{(g)} =0$. The $K_{A}{}^{B}$ terms asymptote to
constants if the warp factor $f(r)$ asymptotes to an exponential, as
we expect it to. The problematic term in ($\ref{eq:ri1}$) is the one
of the form $(m-1)(m-2) / \alpha^{2}$, from which we deduce that the
function $\alpha(r)$ must either asymptote to infinity or a
constant in the limit $r \to \infty$. However, we find that no solution exists to the Einstein equations such that $\alpha \to {\rm const}$ as $r \to \infty$ and
$f(r)$ is a real exponential function. Hence we are left only with
the possibility $\alpha \to \infty$ as $r \to \infty$. This removes
the problematic $\alpha^{-2}$ term in ($\ref{eq:b7}$), and we find
that a valid solution exists in this case, where both $\alpha(r)$
and $f(r)$ are growing exponentials in $r$ (a result found in
Ref. \cite{io1}). Since in this case both warp factors are increasing
functions as $r \to \infty$, it follows that the transverse space
has an infinite volume.

Since we are considering local defects of codimension $m>2$,  the above analysis applies and we will generically
obtain transverse spaces with infinite volume. This infinite volume
is a problem, since we find that gravity cannot be localized on the
brane in such a setup. The reason why we cannot localize gravity is
that the zero mode in the graviton spectrum will not be
normalizable. To avoid (but not solve) this problem, we could follow
Ref. \cite{ben1} and introduce an infra-red cutoff when integrating the zero mode
over the extra dimensions. This cutoff could arise as an interbrane
separation, for example. In doing so, we would then obtain a finite integral over the transverse dimensions, and it would be possible to recover conventional four dimensional gravity. Of course, it would be preferable to obtain
a model where gravity can be localized on the brane without the need
to introduce a cutoff, and it appears that global defects are better
suited to achieve this. Alternatively, we could introduce additional
fields in the bulk in an attempt to remove this problematic
behaviour. However, bulk fields have the highly undesirable property
of inducing a non-trivial Weyl tensor contribution to the field
equations on the brane, and a detailed analysis of the bulk would
have to be performed in such a setup.

\section{\label{sec:2} conclusion}

In this paper we have calculated the evolution equation for the
scale factor $a(t)$ of a thick, codimension $m$, $3$-brane. By
assuming radial symmetry in the bulk, we have found that many of the
Einstein equations approximately admit a first integral, and we have integrated
these equations over the brane thickness to obtain a set of
approximate junction conditions at the surface of the brane
$r=\epsilon$. We then used these junction conditions to write an
equation for the evolution of the brane scale factor, $a(t)$, in
terms of the brane energy momentum tensor. Since we considered a
brane of arbitrary codimension, we were forced to make a large
number of assumptions, which we review;

\begin{itemize}

\item  Derivatives tangential to the brane can be neglected.
Specifically, we assumed that the bulk metric has only a weak
dependence on time, and is static to the level of approximation that
we are working at.

\item We assumed that in the core of the brane the metric may be
approximately written as

\begin{equation}\label{eq:ab1} ds^{2} =  g_{\mu\nu}(x)dx^{\mu}dx^{\nu} -
dr^{2} -
r^{2} d\Omega^{2}_{[m-1]} .\end{equation}

\noindent The boundary conditions that we have used give us this
form of the metric at the centre of the brane, that is at $r=0$, and
we have assumed that the metric can be approximately written as ($\ref{eq:ab1}$)
for $r<\epsilon$.

\item We defined four dimensional quantities as $n$ dimensional
quantities integrated over the brane thickness in the transverse
dimensions. For example, the brane energy momentum tensor
$\hat{T}_{A}{}^{B}$ is given by

\begin{equation} \label{eq:cl1} \hat{T}_{A}{}^{B} \equiv \int \sqrt{\gamma}
d^{m-1}y \int_{0}^{\epsilon} \alpha^{m-1}N A^{n} T_{A}{}^{B} dr .
\end{equation}

This is not an assumption, but rather a definition. However, there
is some ambiguity in defining a four dimensional quantity when
discussing thick braneworlds, and ($\ref{eq:cl1}$) is not unique.

\item We have assumed that a solution to the full $n$-dimensional
Einstein equations exists that respects the above assumptions. It is
important to stress that we have not found a full solution to the
field equations. We expect that a solution exists of the form
postulated, subject to the above assumptions.

\item We have assumed that the brane thickness is time independent, at least to first order. We expect that any matter on the brane will have a backreaction effect on the brane profile, and we have assumed that this effect is negligible.
We have also assumed that this thickness is not as small as any
fundamental lengthscale in the model. By this we mean that if the
brane is too thin, then we could not use our classical arguments (we
would not be able to resolve the thickness of the brane without
appealing to quantum mechanics).

\end{itemize}

Based on these assumptions, we have found the standard cosmological
equation plus quadratic terms in the brane energy density for a
thick brane of arbitrary codimension. We have also found a general
fine tuning condition required to make the effective four
dimensional cosmological constant small in our model. It depends on
the bulk energy momentum tensor, the Ricci scalar of the transverse
dimensions, as well as the brane thickness, tension and transverse
energy momentum components $p_{\rm r}$ and $p_{\rm bk}$.
We assumed that the bulk energy momentum tensor in our model is
simply a constant, and we might expect bulk fields to be present
\cite{gh00}. Introducing new fields into the bulk will affect both the fine tuning relationship and the four dimensional Friedmann equation that we have found. Our final result is
the conservation equation for the brane energy momentum tensor,
($\ref{eq:sww1}$). We find that the standard four dimensional
conservation equation is obtained, but only if
the component $T_{\mu r}^{\rm bulk}|_{r=\epsilon}$ of the bulk
energy momentum tensor is zero, a well known result.

\section*{Acknowledgements}

S. A. would like to thank L. Gutteridge for many interesting
discussions during the preparation of this manuscript. This work was
supported by PPARC.

\appendix

\section*{Appendix: Covariant braneworld equations}

In the appendix, we discuss why we obtain a dark radiation term in the brane Friedmann equation, regardless of the codimension, and thereby produce an alternative derivation of the main results of this paper using tensorial notation. We then discuss the relationship between codimension one and codimension $m$ branes in our setup.

\subsection{\label{sec:f1} Weyl tensor}

From our brane Friedmann equation ($\ref{eq:fr1}$), we see that a
dark radiation term is present, which is a bulk effect contributing
to $H$. We might expect extra terms to be introduced into the
Friedmann equation, since we are considering more than one
transverse dimension. We now consider the Weyl tensor in detail, to
determine the origin of this dark radiation term.

The Weyl tensor $W_{\mu\nu}{}^{\alpha\beta}$ is defined as

\begin{equation}\label{eq:wi37} W_{\mu\nu}{}^{\rho\sigma} = R_{\mu\nu}{}^{\rho\sigma} - {4 \over n-2}
g^{[\rho}{}_{[\mu}R^{\sigma]}{}_{\nu]} + {2 \over (n-1)(n-2)}R
g^{[\rho}{}_{[\mu}g^{\sigma]}{}_{\nu]} .\end{equation}

\noindent A certain contraction of this tensor determines the bulks effect on the brane. To see this, we consider the covariant braneworld equations of
arbitrary codimension, given by Ref. \cite{c1}

\begin{equation}\label{eq:ff200} R_{\mu\nu}^{(p)} = {p-2 \over
n-2}\eta_{\mu}{}^{\rho}\eta_{\nu}{}^{\sigma}R_{\rho\sigma} + {1 \over
n-2} \left( \eta^{\rho\sigma}R_{\rho\sigma} - {p-1 \over
n-1}R\right) \eta_{\mu\nu} + {p-1 \over p^{2}}
\bar{K}^{\sigma}\bar{K}_{\sigma}\eta_{\mu\nu} + {p-2 \over p}
\bar{C}_{\mu\nu}{}^{\sigma}\bar{K}_{\sigma}-
\bar{C}_{\mu}{}^{\rho\sigma}\bar{C}_{\nu\rho\sigma} + W_{\mu\nu},
\end{equation}

\noindent where $\eta_{\mu\nu}$ is a tensor that projects other
tensors tangentially to the brane. The Weyl tensor $W_{\mu\nu}$ is
given by

\begin{equation}\label{eq:w1} W_{\mu\nu} = \eta_{\mu}{}^{\sigma}\eta_{\nu}{}^{\kappa}
\eta_{\tau}{}^{\rho}W_{\rho\sigma}{}^{\tau}{}_{\kappa}, \end{equation}

\noindent and $\bar{C}_{\mu\nu}{}^{\rho}$ by

\begin{equation} \bar{C}_{\mu\nu}{}^{\rho} = \bar{K}_{\mu\nu}{}^{\rho} - {1
\over p} \eta_{\mu\nu} \bar{K}^{\rho} .\end{equation}

The extrinsic curvature $\bar{K}_{\mu\nu}{}^{\rho}$ in this notation
is

\begin{equation}\label{eq:ap1}\bar{K}_{\mu\nu}{}^{\rho} =
\eta_{\nu}{}^{\sigma}\eta_{\mu}{}^{\alpha}\nabla_{\alpha}
\eta_{\sigma}{}^{\rho} \end{equation}

\noindent In this section we use $\bar{K}_{\mu\nu}{}^{\rho}$ as the
extrinsic curvature, as opposed to $K_{AB}$ which has been used in
this paper. $K_{AB}$ is actually a
particular example of the more general $\bar{K}_{\mu\nu}{}^{\rho}$
above. To see this, we write ($\ref{eq:ap1}$) as

\begin{align} \label{eq:ex1} \bar{K}_{\mu\nu}{}^{\rho} &=
\eta_{\nu}{}^{\sigma}\eta_{\mu}{}^{\alpha}\perp^{\rho}{}_{\gamma}\nabla_{\alpha}
\eta_{\sigma}{}^{\gamma} \\ &= \eta_{\nu}{}^{\sigma}\eta_{\mu}{}^{\alpha}
\perp^{\rho}{}_{\gamma}\left(\partial_{\alpha}\eta_{\sigma}{}^{\gamma} +
\Gamma^{\gamma}{}_{\alpha\beta}\eta^{\beta}{}_{\sigma} -
\Gamma^{\beta}{}_{\alpha\sigma}\eta_{\beta}{}^{\gamma} \right).
\end{align}

\noindent Next, we note that in our coordinate system,
$\eta_{\mu\nu} = g_{AB}\delta^{A}{}_{\mu}\delta^{B}{}_{\nu}$,
$\eta_{\mu}{}^{\nu} = \delta_{A}{}^{B}
\delta_{B}{}^{\nu}\delta_{\mu}{}^{A}$. Using this, as well as the
relation $\eta_{\mu}{}^{\nu}\perp_{\alpha}{}^{\mu} = 0$, we find that
the extrinsic curvature can be written as

\begin{align} \bar{K}_{\mu\nu}{}^{\rho} &=
\delta_{\mu}{}^{A}\delta_{\nu}{}^{B} \perp^{\rho}{}_{\gamma}
\Gamma^{\gamma}{}_{AB} \\ &= -{1 \over 2}
\delta_{\mu}{}^{A}\delta_{\nu}{}^{B}
\perp^{\rho\gamma}\partial_{\gamma}g_{AB} .\end{align}

\noindent Now, we can use that fact that due to the symmetry imposed
on our metric, the only non zero orthogonal derivative of $g_{AB}$
is in the radial direction, which means we can write

\begin{equation} \bar{K}_{\mu\nu}{}^{\rho} = {1 \over 2}\delta_{r}{}^{\rho}
\delta_{\mu}{}^{A}\delta_{\nu}{}^{B}
\partial_{r}g_{AB} \end{equation}

\noindent where we used $\perp^{r \gamma} = g^{rr} = -1$. The only
non-zero components of $\bar{K}_{\mu\nu}{}^{\rho}$ are $\rho
= r$, and hence we can drop this index, and write

\begin{equation} \bar{K}_{\mu\nu} ={1 \over 2}
\delta_{\mu}{}^{A}\delta_{\nu}{}^{B}
\partial_{r}g_{AB} .\end{equation}

\noindent Comparing $\bar{K}_{\mu\nu}$ in this section with the
$K_{AB}$ that we have been using;

\begin{equation} K_{AB} = \partial_{r} g_{AB} ,\end{equation}

\noindent we see that they differ by a factor of ${1 \over 2}$,
which we must account for in what follows. In addition, we note that
$\bar{K}_{\mu\nu\rho}$ will be used. This is given by
$\perp_{\rho\alpha}\bar{K}_{\mu\nu}{}^{\alpha}$. Since $g_{rr} = -1$, when we lower the
third index on the extrinsic curvature, we must introduce a factor
of $-1$,

\begin{equation} \bar{K}_{\mu\nu\rho} = - \delta_{\rho}^{r}
\delta_{\mu}{}^{A}\delta_{\nu}{}^{B} K_{AB} .\end{equation}

Returning to our codimension one calculation, equation ($\ref{eq:ff200}$) can be used to calculate the brane
evolution equation ($\ref{eq:fr1}$). Note
the presence of the Weyl tensor in ($\ref{eq:ff200}$); for codimension one objects
it is $W_{\mu\nu}$ which gives the dark radiation term.

In five dimensional, thin braneworld models, the Weyl tensor is
singular at the position of the brane. For this reason,
$W_{\mu\nu}$ in ($\ref{eq:ff200}$) is not evaluated on
the brane. This is the approach that
we will take; we evaluate the Weyl tensor at some $r = \epsilon + \delta \gtrsim \epsilon$ outside the core. We look for a solution to the Field
equations $R_{\mu\nu} = \Lambda g_{\mu\nu}$, since we have a
cosmological constant only in the bulk. A solution to these field
equations is given by

\begin{equation} \label{eq:l1001} ds^{2} =  -h(a) dt^{2} + {da^{2} \over h(a)} + a^{2}
\left[d\chi^{2} + \chi^{2} \left(d\theta^{2} + \sin^{2}\theta d
\phi^{2}\right)\right] + \alpha^{2}_{0}
\gamma_{ab}dy^{a}dy^{b},\end{equation}

\noindent where

\begin{equation}h(a) = {(n-1) \over 4L^{2}}a^{2} - {\alpha \over
a^{2}} ,\end{equation}

\noindent and the constant $\alpha^{2}_{0}$ is given by
$\alpha^{2}_{0} = (m-2) L^{2} / (n-1)$.

From ($\ref{eq:l1001}$) we can now calculate the Weyl tensor
($\ref{eq:wi37}$). The relevant components that contribute to the
brane evolution equations are $W_{DA}{}^{D B}$, where capital Latin
indices run over the $(3+1)$ brane coordinates. To calculate the
Weyl tensor contribution explicitly, we write $R_{DA}{}^{DB} =
R_{A}{}^{B} - R_{aA}{}^{aB} - R_{rA}{}^{rB}$, and use the fact that
for the metric ($\ref{eq:l1001}$), we can set $R_{aA}{}^{aB} = 0$.
Hence, using $R_{\mu\nu} = \Lambda_{\rm n} g_{\mu\nu}$ and $R = n
\Lambda_{\rm n}$, we can write the relevant Weyl tensor components as,

\begin{equation}\label{eq:wkk1} W_{Di}{}^{Dj} = {(4-n) \over L^{2}}\delta_{i}{}^{j} +
{h'  \over 2 a}\delta_{i}{}^{j} = {15-3n \over 4L^{2}}\delta_{i}{}^{j} +
{\alpha \over a^{4}}\delta_{i}{}^{j} ,\end{equation}

\begin{equation}\label{eq:wkk2} W_{D t }{}^{D t} = {(4-n) \over L^{2}} + {h'' \over
2} = {15-3n \over 4L^{2}} - {3 \alpha \over a^{4}} ,\end{equation}

\noindent We note that the standard dark radiation
term is present in ($\ref{eq:wkk1}$) and ($\ref{eq:wkk2}$).

Thus we have confirmed the presence of the dark radiation term in our
setup. To understand why we obtain this term, we return to the
metric ($\ref{eq:l1001}$). We see that we can split this metric into
two components; a five dimensional part given by

\begin{equation} \label{eq:l10010} ds^{2}_{[5]} =  -h(a) dt^{2} + {da^{2} \over h(a)} + a^{2}
\left[d\chi^{2} + \chi^{2} \left(d\theta^{2} + \sin^{2}\theta d
\phi^{2}\right)\right] ,\end{equation}

\noindent which is the standard five dimensional metric considered
in the literature \cite{ma04}, that is the standard five dimensional
Schwarzschild AdS line element, and a second component

\begin{equation}\label{eq:rm1} ds^{2}_{[m-1]} = \alpha^{2}_{0}
\gamma_{ab}dy^{a}dy^{b},\end{equation}

\noindent which is simply pure AdS. This second component is the
$(m-1)$ codimensions. From this split the origin of the dark
radiation term becomes a little clearer; we might expect to obtain a
dark radiation term from the five dimensional part
($\ref{eq:l10010}$) of our metric. The remaining $(m-1)$
codimensions in ($\ref{eq:rm1}$) are pure AdS only, and hence will
contribute only terms like $\sim 1/L^{2}$ to $W_{\mu\nu}$. The fact
that we obtain a dark radiation term in the brane Friedmann equation
is a consequence of our choice of metric ansatz. If we dropped our
assumption of spherical symmetry in the extra dimensions, or
introduced additional bulk fields, then we would obtain more
complicated bulk effects on the brane scale factor. In other words, in this particular model we have over-constrained the bulk, and assumed that it is simply pure AdS away from the brane.

\subsection{Covariant braneworld equations}

Finally, we verify that we can obtain our equations using
the covariant braneworld equations given in, for example, \cite{c1}.
We will see that we can consider our brane as either a codimension
one or codimension $m$ object, and still obtain the same brane
equation. We will explain why this is so at the end of the section.

We begin with the generalized Gauss equation for a $p$-brane of
arbitrary codimension, as found in \cite{c1}. This equation relates
the Ricci tensor of the brane, $R^{(p)}$, to the full
$n$-dimensional Ricci tensor $R_{\mu\nu}$, the extrinsic curvature
$\bar{K}_{\mu\nu}{}^{\rho}$ and the appropriately contracted Weyl
tensor $W_{\mu\nu}$. We will show that we can obtain our brane
Friedmann equation using two approaches. In the first method, we
consider a codimension one object, with the $(m-1)$ `codimensions'
not as transverse dimensions but rather as brane parallel
dimensions. In this approach the radial coordinate $r$ acts as the
orthogonal coordinate, and we find our Friedmann equation arises
from the Gauss equation.

In the second approach, we consider our brane as a codimension $m$
object, and calculate the Weyl tensor and background Ricci tensor in
terms of the extrinsic curvatures $K_{A}^{B}$ and $L_{a}^{b}$. With
this approach, we obtain the same Friedmann equation. We will show
this, and then explain why we obtain the same result regardless of
whether we consider the $(m-1)$ codimensions as brane parallel or
brane orthogonal directions.

In \cite{c1}, the Ricci tensor $R_{\mu\nu}^{(p)}$ and scalar
$R^{(p)}$ of a $p$-brane embedded in an $n$-dimensional background
space have been calculated, and are given by

\begin{equation}\label{eq:ff2} R_{\mu\nu}^{(p)} = {p-2 \over
n-2}\eta_{\mu}{}^{\rho}\eta_{\nu}{}^{\sigma}R_{\rho\sigma} + {1 \over
n-2} \left( \eta^{\rho\sigma}R_{\rho\sigma} - {p-1 \over
n-1}R\right) \eta_{\mu\nu} + {p-1 \over p^{2}}
\bar{K}^{\sigma}\bar{K}_{\sigma}\eta_{\mu\nu} + {p-2 \over p}
\bar{C}_{\mu\nu}{}^{\sigma}\bar{K}_{\sigma}-
\bar{C}_{\mu}{}^{\rho\sigma}\bar{C}_{\nu\rho\sigma} + W_{\mu\nu},
\end{equation}

\noindent where

\begin{equation}\label{eq:w1} W_{\mu\nu} = \eta_{\mu}{}^{\sigma}\eta_{\nu}{}^{\kappa}
\eta_{\tau}{}^{\rho}W_{\rho\sigma}{}^{\tau}{}_{\kappa} ,\end{equation}

\noindent and

\begin{equation}\label{eq:rp1} R^{(p)} = {p-1 \over n-2} \left( 2
\eta^{\rho\sigma}R_{\rho\sigma} - {p \over n-1}R \right) + {p-1
\over p} \bar{K}^{\sigma}\bar{K}_{\sigma} -
\bar{C}_{\lambda\mu}{}^{\nu} \bar{C}^{\lambda\mu}{}_{\nu} + W,
\end{equation}

\noindent where

\begin{equation} \bar{C}_{\mu\nu}{}^{\rho} = \bar{K}_{\mu\nu}{}^{\rho} - {1
\over p} \eta_{\mu\nu} \bar{K}^{\rho} .\end{equation}

\noindent To begin, we consider a codimension one object of
dimension $p = m+3$. If we consider our analysis as describing a
codimension one object, with $r$ being the codimension, then
$R^{(p)} = R^{(g)} + R^{(\gamma)}$, and we can calculate our brane
Friedmann equation from ($\ref{eq:rp1}$).

To do so, we will need to evaluate the
projected Weyl tensor, specifically the trace of ($\ref{eq:w1}$),

\begin{equation}\label{eq:w} W = \eta^{\kappa\sigma}
W_{\rho\sigma}{}^{\tau}{}_{\kappa}\eta^{\rho}{}_{\tau} .\end{equation}

\noindent Remembering that the brane tangential projection
$\eta_{\mu\nu}$ now runs over the standard four dimensions as well
as the $(m-1)$ spherically symmetric dimensions, the relevant
components of ($\ref{eq:w}$) are given by

\begin{equation}\label{eq:wi1} W_{Di}{}^{Dj} = {(4-n) \over L^{2}}\delta_{i}{}^{j} +
{h'  \over 2a}\delta_{i}{}^{j} = -{15-3n \over
4(n-1)}\Lambda_{\rm n}\delta_{i}{}^{j} + {\alpha \over
a^{4}}\delta_{i}{}^{j},
\end{equation}

\begin{equation}\label{eq:wi2} W_{D t }{}^{D t} = {(4-n) \over L^{2}} + {h'' \over
2} = -{15-3n \over 4(n-1)}\Lambda_{\rm n} - {3 \alpha \over a^{4}},
\end{equation}

\begin{equation} \label{eq:wi3} W_{ab}{}^{ab} = {5 (m-1) \over
(m+3)} \Lambda_{\rm n} ,\end{equation}

\begin{equation} \label{eq:wi4} W_{aB}{}^{aB} + W_{Ba}{}^{Ba} =
-{8(m-1) \over m+3}\Lambda_{\rm n} .\end{equation}

\noindent Now taking the trace of ($\ref{eq:wi1}$) and summing over
all of the above Weyl tensor contributions to $W$ gives the result
$W = 0$, as it should be; the trace of the Weyl tensor is zero for
codimension one objects.

Using all of the above in ($\ref{eq:rp1}$), as well as the fact that
$p = m+3$, $n = m+4$, gives us

\begin{equation}\label{eq:dk1} R^{(g)} + R^{(\gamma)} = 2\left(R^{a}{}_{a} +
R^{A}{}_{A}\right) - R_{\alpha}{}^{\alpha} - {(m+2) \over
4(m+3)}\left(K^{2} + L^{2} + 2KL\right) + {1 \over
4}\left(K_{A}{}^{B}K_{B}{}^{A} + L_{a}{}^{b}L_{b}{}^{a}\right) - {1 \over
4(m+3)}\left( K^{2} + L^{2} + 2KL \right) .\end{equation}

\noindent Finally, we use the fact that

\begin{equation} R^{(g)} = -6 \left( {\ddot{a} \over a} +
\left({\dot{a} \over a}\right)^{2} \right) ,\end{equation}

\noindent as well as

\begin{equation}\label{eq:as1} 2R_{r}{}^{r} -  R_{\alpha}{}^{\alpha} =
R_{\alpha}{}^{\alpha} - 2\left( R_{a}{}^{a} + R_{A}{}^{A} \right) = 2
{T_{r}{}^{r} \over M^{N-2}} ,\end{equation}

\noindent to write ($\ref{eq:dk1}$) as

\begin{equation}\label{eq:jkb20} -6 \left( {\ddot{a} \over a} +
\left({\dot{a} \over a}\right)^{2} \right) = - 2{T_{r}{}^{r} \over
M^{n-2}} - R^{(\gamma)} - {1 \over 4} \left( K^{2} + L^{2} + 2
KL\right) + {1 \over 4} \left( K_{C}{}^{D}K_{D}{}^{C} +
L_{a}{}^{b}L_{b}{}^{a}\right) .\end{equation}

\noindent Next, we consider the field equations 'junction conditions' for our thick brane approach. As before, we must integrate them over the brane orthogonal coordinates. However, since we are now simply considering a codimension one object, we only integrate over the $r$ coordinate. Doing so, we obtain

\begin{equation}\label{eq:jkb17} K_{i}{}^{j}|_{r=\epsilon} = { M_{\rm b}^{m-1}
\over M^{n-2}}\left(\tilde{T}_{i}{}^{j} -
\delta_{i}{}^{j}{\tilde{T} \over n-2}\right) ,\end{equation}

\begin{equation}\label{eq:jkb18} K_{t}{}^{t}|_{r=\epsilon} = { M_{\rm b}^{m-1}
\over M^{n-2}}\left(\tilde{T}_{t}{}^{t} -
\delta_{t}{}^{t}{\tilde{T} \over n-2}\right) ,\end{equation}

\begin{equation}\label{eq:jkb19} L_{a}{}^{b}|_{r=\epsilon} = { M_{\rm b}^{m-1}
\over M^{n-2}}\left(\tilde{T}_{a}{}^{b} -
\delta_{a}{}^{b}{\tilde{T} \over n-2}\right) -{2 \over
\epsilon}\delta_{a}{}^{b},\end{equation}

The final step is to use ($\ref{eq:jkb17}$-$\ref{eq:jkb19}$) to write
($\ref{eq:jkb20}$) as

\begin{equation}
3\left({\ddot{a}\over a}+ \left({\dot{a}
\over a}\right)^{2}\right)+ {M_{\rm b}^{2m-2} \over 8 (n-2)^{2}M^{2n-4}}
\left((n-2)^{2}\tilde{T}_{A}{}^{B}\tilde{T}_{B}{}^{A} +
B_{m}(\tilde{T}_{A}{}^{A})^{2} +
(n-2)^{2}\tilde{T}_{a}{}^{b}\tilde{T}_{b}{}^{a} +
B_{m}(\tilde{T}_{a}{}^{a})^{2}\right)
\end{equation} \begin{equation*}+ {M_{\rm b}^{2m-2} \over 8 (n-2)^{2}M^{2n-4}}\left(2B_{m}\left(  \tilde{T}_{a}{}^{a} +
\tilde{T}_{r}{}^{r}\right) \tilde{T}_{A}{}^{A} + 2B_{m}
\tilde{T}_{a}{}^{a}\tilde{T}_{r}{}^{r} +
C_{m}(\tilde{T}_{r}{}^{r})^{2}\right) - {M_{\rm b}^{m-1} \over 2
\epsilon M^{n-2}}\tilde{T}_{a}{}^{a} \end{equation*} \begin{equation*}- {(m-1)M_{\rm b}^{m-1} \over 2\epsilon M^{n-2}}
\tilde{T}^{r}{}_{r}- { (m-1)(m-2)\over \epsilon^{2}} = -{\Lambda
\over M^{n-2}}.\end{equation*}

\noindent which is the same equation as ($\ref{eq:b25}$) for a codimension $m$ brane. Note that since we have assumed that
our space contains a codimension one object, the trace of the Weyl
tensor vanishes. However, $W_{\mu\nu}$ does not vanish; as we have
shown above we obtain a dark radiation like term.

It may seem quite unnatural that we obtain the same Friedmann
equation whether we consider a codimension one or $m$ object. The
reason we do so is because we have assumed spherical symmetry. In
our paper, we have considered surfaces of constant $r = \epsilon$
only; since the other codimensions are spherically symmetric we do
not have to set them to a particular fixed value (since our final
result will not depend on our choice). Hence the $(m-1)$ spherically
symmetric dimensions could equally well be brane tangential or brane
orthogonal coordinates. The only difference between the two would be
the form of the $(m-1)$ components of the energy momentum tensor
$\tilde{T}_{a}{}^{b}$, which would be small for a codimension $m$ brane, but potentially large for a codimension one brane.

We now proceed with our second approach, that is to consider our
brane as a four-dimensional object of codimension $m$. Now, we will
no longer have the trivial result $W=0$, however we find that we
still obtain the same equation relating the brane scale factor
$a(t)$ with $\tilde{T}_{A}{}^{B}$, $\tilde{T}_{r}{}^{r}$ and
$\tilde{T}_{a}{}^{b}$.

To proceed, we consider ($\ref{eq:rp1}$) again. Now we have $p=4$
and $R^{(p)} = R^{(g)}$. We begin by calculating the projected Weyl
tensor, using ($\ref{eq:wi37}$). The relevant components in our
coordinate system are $W_{AB}{}^{AB}$, which are explicitly

\begin{equation}\label{eq:wk1} W_{AB}{}^{AB} = R_{AB}{}^{AB} - {6 \over
(n-2)}R^{A}{}_{A} + {12 \over (n-1)(n-2)}R_{\alpha}{}^{\alpha}.
\end{equation}

\noindent Using ($\ref{eq:wk1}$) in ($\ref{eq:rp1}$), we find that
the brane Ricci tensor $R^{p}$ may be written as

\begin{equation}\label{eq:as2} R^{(p)} = R_{AB}{}^{AB} - {3 \over 16} K^{2} + {1
\over 4} K^{A}_{B}K^{B}_{A} - {1 \over 16} K^{2} .\end{equation}

\noindent Next, we use the relation ($\ref{eq:as1}$), as well as

\begin{equation} R_{AB}{}^{AB} = R_{A}{}^{A} - R_{aA}{}^{aA} -
R_{rA}{}^{rA} \end{equation}

\noindent to write ($\ref{eq:as2}$) as

\begin{align}  R^{(p)} &= R^{\alpha}{}_{\alpha} -
2R^{a}{}_{a} - R_{A}{}^{A} - R_{aA}{}^{aA} - R_{rA}{}^{rA} + {1 \over 4}
K^{A}{}_{B}K^{B}{}_{A} - {1 \over 4} K^{2} - 2{T_{r}{}^{r} \over M^{N-2}}
\\ \label{eq:as3} &= R_{r}{}^{r} - R_{a}{}^{a} - R_{aA}{}^{aA} - R_{rA}{}^{rA} + {1 \over 4}
K^{A}{}_{B}K^{B}{}_{A} - {1 \over 4} K^{2} - 2{T_{r}{}^{r} \over M^{n-2}}
\end{align}

\noindent we now evaluate the terms in ($\ref{eq:as3}$). We find

\begin{equation} R_{r}{}^{r} - R_{a}{}^{a} ={1 \over 2}
K' + {1 \over 4}\left( K_{A}{}^{B}K_{B}{}^{A} + L_{a}{}^{b}L_{b}{}^{a}
\right) - R^{(\gamma)} - {1 \over 4}L^{2} - {1 \over 4}KL ,
\end{equation}

\begin{equation} -R_{aA}{}^{aA} = -{1 \over 4}KL ,\end{equation}

\begin{equation} - R_{rA}{}^{rA} = -{1 \over 2}K' - {1 \over 4}
K_{A}{}^{B}K_{B}{}^{A} .\end{equation}

\noindent Using these in ($\ref{eq:as3}$) gives us

\begin{equation} \label{eq:as4} R^{(p)} = {1 \over
4}\left(K_{A}{}^{B}K_{B}{}^{A} + L_{a}{}^{b}L_{b}{}^{a}\right) - {1 \over 4}
\left(K^{2} + L^{2}\right) - {1 \over 2}KL - 2{T_{r}{}^{r} \over
M^{n-2}} - R^{(\gamma)} .\end{equation}

\noindent Once again, if the then proceed to write $K_{A}{}^{B}$ and
$L_{a}{}^{b}$ in terms of $\tilde{T}_{\mu}{}^{\nu}$, we would find
the same equation as we found in both the previous sections and
above (when we considered the brane as a codimension one object). We
stress that we obtain the same result because of the symmetry
imposed on the $(m-1)$ codimensions. This means that our choice of
the angular coordinates will not affect our Ricci scalar $R^{(p)}$,
and hence we can consider the spherically symmetric coordinates
either as being brane orthogonal or brane tangential.

Finally, we note that our result that codimension one and $m$
branes are equivalent for a metric such as ours is only valid for a
thick brane of codimension one or $m$. If we consider thin branes,
then in the codimension $m$ case we would obtain $\delta$-function
singularities in $K_{A}{}^{B}$ and hence $R^{(p)}$, whereas for
codimension one branes there would be no divergent behaviour. The
singular behaviour is removed in the codimension $m$ case since we
have smeared the brane energy momentum tensor over a finite region
of space.

The brane Friedmann equation that we have derived is written in
terms of the brane energy momentum tensor, which has the standard
four dimensional components $\tilde{T}_{AB}$, and in addition non
zero components in the $(a,b)$ and $(r,r)$ directions,
$\tilde{T}_{ab}$ and $\tilde{T}_{rr}$. Above, we have shown that we
can write down an evolution equation for $a(t)$ in terms of
$\tilde{T}_{AB}$, $\tilde{T}_{ab}$ and $\tilde{T}_{rr}$, regardless
of whether the object is codimension one or $m$. The only difference
between the two setups will be the form of the $\tilde{T}_{ab}$
components of the brane energy momentum tensor. When we have $m$ codimensions, we
expect that $\tilde{T}_{ab}$ will be some small quantity, which
asymptotes to zero in the thin brane limit. For a codimension one
brane however, the $\tilde{T}_{ab}$ will have the same form as the
$\tilde{T}_{AB}$ components. This is the principle difference
between the two treatments. However in this paper we do not
explicitly specify the brane energy momentum tensor.

\end{document}